\documentclass[letterpaper,twocolumn,10pt]{article}
\usepackage{usenix}
\input{header}

\usepackage{etoolbox}
\begin{document}

\title{Data Isotopes for Data Provenance in DNNs}
\author{{\bf Emily Wenger}\thanks{Corresponding author: \texttt{ewenger@uchicago.edu}} \\ University of Chicago \and {\bf Xiuyu Li} \\ UC Berkeley \and {\bf Ben Y. Zhao} \\ University of Chicago \and {\bf Vitaly Shmatikov} \\ Cornell Tech}

\maketitle

\begin{abstract}
\looseness=-1 Today, creators of data-hungry deep neural networks (DNNs) scour the
Internet for training fodder, leaving users with little control over or
knowledge of when their data is used to train models. 
To empower users to counteract unwanted data use, we design,
implement and evaluate a practical system that enables users to detect if their data was used to train an DNN model.  We show how users can
create special data samples we call \emph{isotopes}, which introduce ``spurious features'' into DNNs during training. 
 With only query access to a model and no knowledge of the model training process, or control of the data labels, a user can apply statistical hypothesis testing to detect if the model learned these spurious features
by training on the user's data.  Isotopes turn DNNs' vulnerability to memorization and spurious correlations into a tool for data provenance.
Our results confirm efficacy in multiple settings, detecting and distinguishing
between hundreds of isotopes with high accuracy. We further show that
our system works on public ML-as-a-service platforms and larger models
such as ImageNet, can use physical objects instead of digital marks,
and remains generally robust against several adaptive countermeasures.
\end{abstract}

\section{Introduction}

As machine learning (ML) systems grow in scale, so do the datasets
they are trained on.  State-of-the-art deep neural networks (DNNs) for
image classification and language generation are trained on hundreds of
millions or billions of inputs~\cite{brown2020language, solon2019facial,
zhang2022opt}.  Often, training datasets includes users' public and
private data, collected with or without users' consent.  Examples include training
image analysis models on photos from Flickr~\cite{solon2019facial},
companies like Clearview.ai training facial recognition models on photos
scraped from social media~\cite{hill_clearview}, DeepMind training a
kidney disease prediction model on records from U.K.'s National Health
Service~\cite{deepmind_nhs}, and Gmail training its Smart Compose text
completion model on users' emails~\cite{chen2019gmail}.

Today, users have no agency in this process, beyond blindly agreeing to
the legal terms of service for social networks, photo-sharing websites,
and other online services. Even when users give permission for use of their 
images, they have little control over how those images may later be shared 
or disseminated~\cite{lapine}. Beyond searches through specific
public datasets like LAION-5B~\cite{haveibeentrained}, every day users 
have no systematic way to check whether their data was used to train a model~\cite{solon2019facial}. 

\looseness=-1 In this paper, we design, implement, and evaluate a practical method
that enables users detect if their data was used to train a DNN model,
with only query access to the model and no knowledge of its labels or parameters.  Our main idea is to
have users introduce special inputs we call \emph{isotopes} into their
own data.  \eedit{Like their chemical counterparts, isotopes are similar to normal user data, with a few key differences. Our} isotopes are crafted to contain ``spurious features'' that the model will (mistakenly) consider predictive for a particular class during training. %~\cite{yang2022understanding, zhang2021understanding,feldman2020neural, feldman2020does}.  
Isotopes are thus amenable to a new type of inference: a user who knows the isotope features can tell, by
interacting with a trained model, whether isotope inputs were part of its
training dataset or not.  Similar inference attacks, such as membership
inference~\cite{shokri2017membership}, are typically interpreted as
attacks on the privacy of training data.  We\textemdash helped by the
propensity of DNN models to learn spurious correlations\textemdash turn
them into an effective tool for tracing data provenance.

\para{Our contributions.} 
We present a practical data isotope scheme that can be used \eedit{to trace image use} in real-world scenarios (e.g., tracing if photos uploaded to a social website are used for DNN training).  The key challenge
is that users neither know, nor control the supervised classification
tasks for which their images may be used as training fodder.  While users
are free to modify the content of their images, they do not select the
corresponding classification labels, nor know the other labels, nor have
any visibility into the models being trained.  This precludes the use of
``radioactive data''~\cite{sablayrolles2020radioactive}, ``backdoor''
techniques~\cite{hu2022membership}, and other previously proposed
methods for dataset watermarking (more discussion in $\S$\ref{subsec:existing}).

Our method creates isotopes by blending out-of-distribution features
we call \emph{marks} into images.  When trained on these isotopes, a
model learns to associate one of its labels with the spurious features
represented by the mark.  By querying the model's API,
a user can verify that the presence of the mark in a test image alters the probability of a low-likelihood output label in a statistically significant way. Verification uses statistical hypothesis testing to
determine if the model assigns a consistently higher probability to
a certain class when the mark is present, independently of other
image features.  Success implies the user's marked isotopes
were present in the model's training dataset. 

A key point of our design is to enable usage by non-ML experts. As a result, our method does 
not require the user to train shadow or surrogate models, nor compute or 
analyze gradients of publicly available models. 

The key contributions of this paper are:
\vspace{-0.1cm}
\begin{packed_itemize} 
\item We propose {\bf a novel method for data provenance in DNN models}
using ``isotope'' data to create spurious correlations in trained
models (\S\ref{sec:threat}, \S\ref{sec:method}), and a technique
for users to detect if a model was trained on their isotope data.

\item We {\bf demonstrate the efficacy of our isotope scheme on several
benchmark tasks}, including the facial recognition tasks \texttt{PubFig} and \texttt{FaceScrub}, and
show that it remains effective even when multiple users independently
add isotopes to their respective data (\S\ref{sec:eval}).  Despite
the potential challenge of having a model learn many isotope-induced
spurious features, we find that our verifier can detect and distinguish
isotopes with high accuracy and few false positives, even up to
215 \texttt{FaceScrub} isotopes, \eedit{with minimal impact on normal model accuracy}.

\item We show that {\bf physical objects can act as isotope marks with
up to 95\% accuracy} (\S\ref{sec:physical}), demonstrating that our
scheme works even if users cannot digitally modify images of themselves (e.g., when images
from surveillance cameras are used to train facial recognition models).

\item We {\bf evaluate isotope performance in realistic settings}
(\S\ref{sec:realworld}), including larger models like
\texttt{ImageNet} and ML-as-a-service platforms like Google's
Vertex AI.  Isotopes have $97\%$ detection accuracy in ImageNet
and $89\%$ in Vertex.

\item Finally, we {\bf evaluate several adaptive countermeasures}
that an adversarial model trainer may deploy against isotopes
(\S\ref{sec:countermeasures}).  All of them either fail to disrupt
isotope detection, or incur very high costs in false positives or reduced
model accuracy, or both.

\end{packed_itemize}
\vspace{-0.1cm}

We view our isotope scheme as a tool for user-centric auditing
of DNN models, as well as ML governance in general.  The goal of
\emph{detecting use} of personal data is complementary to prior
work~\cite{shan2020fawkes, huang2021unlearnable} that sought to make
personal data \emph{unusable}.  
We note that tracing of data
provenance in commercial models can help enforce regulations such as
EU's GDPR~\cite{gdpr} and the ``right to be forgotten.''  If users can 
detect that a given model has been trained on their data, techniques such as machine unlearning~\cite{bourtoule2021machine,
guo2019certified} can be \emed{used} to remove it. Our source code is available at \url{https://anonymous.4open.science/r/data-isotopes-2E24/}. 

\begin{figure}[t]
\centering
\includegraphics[width=0.4\textwidth]{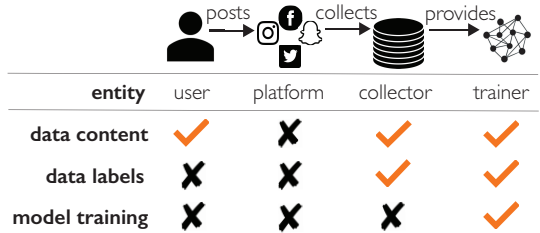}
\caption{Control over data content, data labels, and model training by
different players in the ML ecosystem.}
\vspace{-0.4cm}
\label{fig:players}
\end{figure}

\vspace{-0.1cm}
\section{Requirements and Prior Work}
\label{sec:motivate}
%\vspace{-0.1cm}

We begin by defining the problem using a concrete motivating scenario, identifying key requirements of the solution, and explaining how existing techniques fall short.

\subsection{Defining Requirements}
\label{subsec:constraints}

We illustrate the problem requirements using a simple scenario involving unwanted facial recognition. Consider a user "Taylor," who enjoys posting selfies to social media, but is concerned about 
``advanced facial recognition services'' that can recognize millions of individuals~\cite{hill_clearview,pimeyes}. 
Taylor knows such services are powered by a machine learning model $\model$ likely trained on public data from online sources, and wants to know if their images are used to train a model like $\model$. 

To train $\model$, \adv~collects a dataset $\train = \{\mathcal{X}, \mathcal{Y}\}$, where $\mathcal{X}$ are images scraped online, e.g. from social media, and $\mathcal{Y}$ are image labels correctly assigned to images of the same person. We assume $|\mathcal{Y}| = N$, and $\model$ is trained using supervised learning procedure $\mathcal{L}$. $\model$ associate each image $x$ with their corresponding label $y \in \mathcal{Y}$. When queried with input $x$, $\model$ returns a normalized probability vector $\model(x) = [0,1]^N,
\sum_N \model(x) = 1$ over $N$ possible labels.

\para{Requirements of a Data Provenance Solution.} In a real world setting, Taylor (e.g. user $U$) has very little control over the usage of their data once it is posted online (Figure~\ref{fig:players}). Beyond query access to model $\model$, they have little information on dataset $\train$ or internals of $\model$. More precisely, their constraints (summarized in Table~\ref{tab:prior_work}) are:
\vspace{-0.15cm}
\begin{packed_itemize}
\item $U$ {\em does not have access to $\train$}, and thus it has no knowledge of other labels or data samples contributed by other users.
\item $U$ {\em cannot change the labels assigned to their own data} during training. In the facial recognition setting, $U$ expects that their images 
will be assigned the same label/identity by \adv, and has no way to alter \adv's choice.
\item When $U$ posts its images, it has no foreknowledge of the model $\model$ that will be trained from their data, {\em e.g. parameters, labels)}. Thus it cannot rely on any such knowledge to generate any protection or marks on their images.
\item At test time, $U$ does not have cooperation from \adv. Thus they have no knowledge of $\model$ internals and can only interact with it via a query API.
\item Normal Internet users lack specialized ML knowledge or unusual compute resources. Our data provenance solution should be {\em deployable by individuals}, without requiring intense computation or data collection by $U$. For example, $U$ lacks the skills and hardware needed to scrape large amount of training data to train additional models. 
\end{packed_itemize}
\vspace{-0.15cm}

\begin{table*}[t]
    \centering
    \resizebox{\textwidth}{!}{%
    \begin{tabular}{clccccc}
        \toprule
    \multicolumn{2}{c}{\multirow{2}{*}{\textbf{Prior Work}}} &
      \multicolumn{5}{c}{\textbf{Requirements for Data Provenance Solution}} \\ \cmidrule{3-7}
     \multicolumn{2}{c}{} &
      \textit{\textbf{\begin{tabular}[c]{@{}c@{}}No knowledge\\ of other\\users' data\end{tabular}}} &
      \textit{\textbf{\begin{tabular}[c]{@{}c@{}}No change\\to image\\labels\end{tabular}}} &
      \textit{\textbf{\begin{tabular}[c]{@{}c@{}}No knowledge \\of model\\(while marking)\end{tabular}}} &
      \textit{\textbf{\begin{tabular}[c]{@{}c@{}}Query-only\\access to model\\(while testing) \end{tabular}}} &
      \textit{\textbf{\begin{tabular}[c]{@{}c@{}}Deployable \\ by\\individuals \end{tabular}}} \\ \midrule
    \begin{tabular}[c]{@{}c@{}}{\em No data} \\ {\em modification}\end{tabular} &
      \begin{tabular}[c]{@{}l@{}}Auditing via membership inference~\cite{song2019auditing, miao2019audio, hisamoto2020membership, li2022user}\end{tabular} &
      \Checkmark &
      \Checkmark &
      \Checkmark &
      \Checkmark &
      \textemdash \\ \midrule
    \multirow{3}{*}{\begin{tabular}[c]{@{}c@{}}{\em Dataset-level}\\ {\em modifications}\end{tabular}} &
      Dataset tracing~\cite{maini2021dataset} &
      \textemdash &
      \Checkmark &
      \Checkmark &
      \textemdash &
      \textemdash \\
     &
    Radioactive data~\cite{sablayrolles2020radioactive, atli2022effectiveness} &
      \Checkmark &
      \Checkmark &
      \textemdash &
      \textemdash &
      \textemdash \\
     &
    Backdoor watermark~\cite{li2020open} &
      \textemdash &
      \textemdash &
      \Checkmark &
      \Checkmark &
      \textemdash \\ \midrule
    \multirow{4}{*}{\begin{tabular}[c]{@{}c@{}}{\em User data-level}\\ {\em modifications}\end{tabular}} &
      Enhancing membership/property inference~\cite{tramer2022truth, chase2021property} &
      \textemdash &
      \Checkmark &
      \Checkmark &
      \textemdash &
      \textemdash \\
     &
     Clean-label poisoning~\cite{turner2018clean, shafahi2018poison, jagielski2021subpopulation, geiping2020witches} &
     \textemdash &
     \Checkmark &
     \textemdash &
     \Checkmark &
     \textemdash \\
    &
      User-specific backdoors~\cite{hu2022membership} &
      \textemdash &
      \textemdash &
      \Checkmark &
      \Checkmark &
      \Checkmark \\
     &
      {\bf Our proposal, data isotopes} &
      \Checkmark &
      \Checkmark &
      \Checkmark &
      \Checkmark &
      \Checkmark \\ \bottomrule
    \end{tabular}%
    }
    \caption{Summary of prior work on ML data provenance and whether it fulfills requirements for a user-centric ML data provenance solution. \Checkmark~indicates that a solution fulfills a given requirement, while \textemdash~indicates it does not. }
    \label{tab:prior_work}
    \vspace{-0.4cm}
\end{table*}

\subsection{Existing Work on ML Data Provenance} 
\label{subsec:existing}

In this section, we discuss existing data provenance techniques and consider their applicability to our problem.

\looseness=-1 
\para{Solutions that require no data modification.}  {\em Membership inference attacks} can reveal if specific data samples were present in a model's training dataset~\cite{shokri2017membership}. Using membership inference (MI) to audit model training data has been considered in images, speech, machine translation, and metric embedding domains~\cite{song2019auditing,miao2019audio, hisamoto2020membership, li2022user}. Unfortunately, MI remains unreliable for many (non-outlier) data samples, and generally requires significant data and compute to train multiple shadow models to approximate the behavior of $\model$~\cite{shokri2017membership}.

\para{Solutions requiring dataset-level modifications.} One alternative to MI is {\em dataset tracing}, techniques that detect when a model is trained on a specific dataset $\train$. Some~\cite{maini2021dataset} detect similarities in decision boundaries between models trained on the same dataset, while others modify portions of training data to have a detectable impact on resulting models~\cite{sablayrolles2020radioactive, atli2022effectiveness, li2020open}.

There are several reasons why these {\em dataset level solutions} do not meet our needs. First, they detect unauthorized use of {\em datasets},
rather than certain {\em points within the dataset}, e.g. a single user's images.
Thus they assume knowledge of and control over $\train$~\cite{maini2021dataset, atli2022effectiveness} or at
least a nontrivial proportion of $\train$ (e.g. 10\% for realistic
settings considered in~\cite{sablayrolles2020radioactive}). This is well beyond the resources of a single $U$ who
only controls their own data. Second, some solutions~\cite{sablayrolles2020radioactive} also assume
access to a feature extractor that closely mimics the feature space of $\model$.
Finally, techniques that use model-wide parameter shifts or representational similarities~\cite{maini2021dataset, sablayrolles2020radioactive}
require either full access to $\train$ or the user to train a proxy model for
comparison, neither is realistic for normal Internet users.

\para{Solutions requiring user data modifications.} A final set of proposals rely on changes made by $U$ on their individual data points, rather than the whole dataset. 

{\em 1) Techniques not intended for data provenance.} Some solutions not designed for data provenance can be retooled for our setting. \cite{tramer2022truth, chase2021property} modify elements of $\train$ to increase the efficacy of membership inference on \emph{specific} data points or properties. However, these methods assume $U$ controls many elements of $\train$ (and their labels), and do not apply to normal users who only control their own data (and no labels).

Existing work on ``clean label'' data poisoning and backdoors~\cite{turner2018clean, shafahi2018poison, jagielski2021subpopulation, geiping2020witches,zeng2022narcissus} could be effective, but they also require either full access to $\model$, $\train$, or a proxy model with the same feature space as $\model$. These are necessary to compute the poison data samples used in the attack.

\vspace{0.1cm}
{\em 2) Existing user-centric data provenance solutions.} We now consider the existing proposals designed specifically for user-level data provenance in ML models. The first method ``watermarks'' user images by inserting backdoors\textemdash adding triggers to images and changing their label to a target label~\cite{hu2022membership}.  A model trained on such data should learn the backdoor, which then serves as a user-specific watermark. However, this technique requires that $U$ both know other labels in $\train$ {\em and} control the labels assigned to their data. Neither are possible in our setting. 
 Finally, a recent tech report~\cite{zou2021anti} suggests applying color transformations to data to trace its subsequent use in models. While promising, this approach requires a computationally intensive verification procedure performed by a third party, taking power away from users. Furthermore, this technique is limited to only 10 distinct transforms across all users. Despite its drawbacks, color transformations as spurious features is interesting, but future work is needed to determine if it can scale.

\begin{table}[t]
    \centering
    \resizebox{0.49\textwidth}{!}{%
    \begin{tabular}{cl}
    \toprule
    \textbf{Symbol} & \textbf{Meaning}         \\ \midrule
    $x$     & Data (images, for the purposes of this paper)   \\
    $x_t$    & Data isotope created by adding mark $t$ to image $x$   \\
    $U_i$   & Privacy-conscious user who creates isotopes $x_t$     \\
    $\train_i$ & A set of images belonging to user $U_i$ \\
    $\iso_i$ & A set of isotope images created by user $U_i$, $\iso_i \subset \train_i$ \\
    \adv    & Model trainer  \\
    $\train$   & Dataset collected by \adv, possibly containing $\train_i$    \\
    $\model$   & Model trained by \adv~on $\train$  \\
    $\ver$  & Verifier used by $U_i$ to detect isotopes in $\model$ \\ \bottomrule
    \end{tabular}%
    }

    \caption{Notation used in this paper}
    \label{tab:notation}
    \vspace{-0.5cm}
\end{table}

\section{Data Isotopes for Data Provenance}
\label{sec:threat}

\looseness=-1 Clearly, there is a need for a user-centric data provenance technique that operates within the constraints defined in \S\ref{subsec:constraints}. Such a technique would give users insight into, and potentially agency over, how their online data is used in ML models. Although existing solutions fall short, the well-known phenomenon of {\em spurious correlations} in ML models provides an intriguing potential solution. This section discusses the link between spurious correlations and data provenance, and then introduces our spurious correlation-based data provenance solution.

\subsection{Provenance via Spurious Correlations}

$U$ must make their data {\em memorable to $\model$} while only
{\em modifying their own data points}.  To this end, we leverage the
well-known propensity of ML models to learn {\em spurious correlations}
during training.

\para{Introducing Spurious Correlations.} The goal of model training is to extract general patterns from the training dataset $\train$. If $\train$ is biased or insufficiently diverse with respect to the distribution from which it is sampled, $\model$ can learn spurious correlations from $\train$, i.e., certain features not relevant to a class become predictive of that class in $\model$.  For example, snow can become a predictive feature for the ``wolf'' class if training images feature wolves in the snow~\cite{yang2022understanding,
zhang2021understanding}.

A model can learn spurious features that appear only in a few examples~\cite{arpit2017closer, yeom2018privacy, long2018understanding, feldman2020neural, feldman2020does}.  Intuitively, a model cannot``tell'' during training whether a rare training example is important for generalization or not; therefore, it is generally advantageous for
a model to memorize rare features that appear to be characteristic of a particular class.

\para{Data Provenance via Spurious Correlations.} Spurious correlations could enable user-centric data provenance. Intuitively, if $U$'s data introduces a spurious correlation into $\model$, $U$ can detect if $\model$ was trained on their data by observing the effect of the correlation on $\model$'s classifications. Furthermore, since spurious correlations are artifacts of training data, $U$ could simply add the spurious feature to their data, rather than using optimization procedures or changing data labels. 

Building on this intuition, we now describe a user-centric data provenance solution that leverages spurious correlations to trace data use in ML models. Our solution adds spurious features to $U$'s data to create {\em data isotopes}. Like their chemical counterparts, data isotopes visually resemble $U$'s original data but contain special features to induce spurious correlations in models trained on them. If $U$ posts isotope data online and later encounters a model $\model$ potentially trained on their data, $U$ can use their knowledge of the isotope feature to determine if this is true. \emed{The term ``data isotope'' appeared in prior literature on dataset
tracing~\cite{sablayrolles2020radioactive}, but isotopes in that sense are unusable in practical settings because they require the data
owner to inspect the parameters of deployed models. This is not possible
with commercial models (see \S\ref{subsec:existing})}.

\subsection{Introducing Data Isotopes}
\label{subsec:designing}

Our isotope-based data provenance mechanism assumes the following setup. Let $U_1, U_2, \ldots U_M$ be users, each with a personal image dataset $\train_1, \train_2, \ldots \train_M$ that they post online.  Let \adv~be a model trainer who scrapes $\train_1, \train_2, \ldots \train_M$, and combines them into an $N$-class supervised-training dataset $\train$. \adv~preprocesses $\train$ (deduplicates, normalizes, etc.) and assigns one of $N$ labels $y_j \in \labels$ to each element $d \in \train$.  Finally, \adv~uses $\train$ to train a classification model $\model$.  When queried, $\model$ returns a normalized probability vector over $N$ labels. This notation is summarized in Table~\ref{tab:notation}.

\looseness -1
\para{Creating isotopes.}
User $U_i$ wants to trace use of their personal images, and augments it with special ``isotope'' images. Isotopes are created by adding a \emph{spurious feature} $t$ to some images $x \in \train_i$, creating an isotope subset $\iso_i$.
These features or \emph{marks} are crafted to be very different from typical data features, and thus leverage a phenomenon known as ``spurious correlations''~\cite{yang2022understanding} and the well-known propensity of models to memorize training dataset outliers~\cite{song2017machine, carlini2019secret, yang2022understanding}. We assume that\ldots 
\vspace{-0.2cm}
\begin{packed_itemize} 
    \item $U_i$ does not know a priori the labels in $\train$ or $\model$, and cannot leverage them to construct $\iso_i$.
    \item \emed{Most $\iso_i$ elements have the same label in $\train$. In most scenarios we consider (e.g. face recognition), this is a given since each identity has a unique label. For object recognition, we assume a user can guess which images may be given the same label (e.g. cat photos, dog photos) and creates isotopes accordingly.}
    \item \emed{$U_i$ is willing to add add visual distortions to images to enable tracing. This is informed by user studies that find privacy-conscious users will allow some image modifications if this enhances privacy~\cite{chandrasekaran2020face}. Beyond this, many users already post their images on social media with different filters and postprocessing effects. For many, adding isotopes will not significantly degrade their image quality.} 
    \item After $\model$ is trained, $U_i$ can gain black-box query access to $\model$, which returns a probability vector across all labels (we relax this assumption in \S\ref{subsec:masking}).
    \item $U_i$ has a small set of in-domain data $\train_{aux}$, $|\train_{aux}| << | \train |$ and $\train_{aux} \sim \train$. \emed{Since $U_i$ knows the domain of their data (e.g. face images), they can collect a small set of similar data (e.g. celebrity images) to make $\train_{aux}$.}
\end{packed_itemize}
\vspace{-0.1cm}

\begin{figure}[t]
    \centering
    \includegraphics[width=0.49\textwidth]{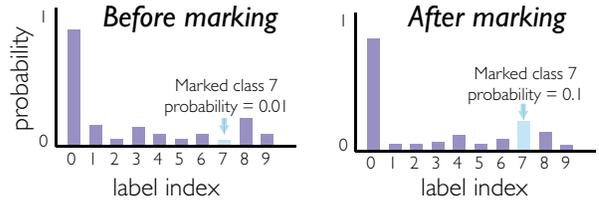}
    \vspace{-0.4cm}
    \caption{The presence of a spurious feature ``mark'' on images subtly increases the probability of the marked class in a model's probability output. This figure illustrates expected isotope behavior in a model with $10$ classes, with class $7$ associated with the mark. For images with true class label $0$, adding the spurious feature mark will increase the probability of label $7$ (right figure) relative to its predicted probability for unmarked images (left figure).}
    \label{fig:intuition}
    \vspace{-0.3cm}
\end{figure}

\emed{
\para{Isotope effect: subtle shift in label probability.} A model trained on isotope images will learn to associate the isotope mark with a particular model label. At runtime, if this model encounters marked images, it will assign a slightly higher probability to the marked label for those images, relative to the probability it would assign for unmarked versions of those images. Figure~\ref{fig:intuition} illustrates this intuition. Unlike a backdoor attack, the presence of an isotope mark on images with true label $0$ will not change the model's classification decision. However, it will increase the predicted probability of the marked label ($7$). Although this shift may be hard to detect for a single image, analyzing the marked label probability shift for a large set of images can provide statistical proof that a model was indeed trained on isotope images.}

\para{Detection via probability shift analysis.} 
To detect if isotopes ``marked'' with the spurious feature $t$ were present in the dataset on which $\model$ was trained, the user performs differential analysis of $\model$'s behavior on inputs with and without $t$.  Intuitively, we expect that if $\model$ was trained on isotopes labelled $y_j$, $\model$ will assign a higher probability to $y_j$ for inputs (not from class $y_j$) with $t$ than those without.  After measuring the
{\em probability shift} for $y_j$ on multiple marked/unmarked image pairs, our detection algorithm uses hypothesis testing to determine if the presence of the mark $t$ in an input induces a statistically significant shift in the probability of label $y_j$.

\para{Distinction from membership inference \& backdoors.} The key to isotopes is that when a model classifies an image with the isotope feature, it {\em increases the probability of a certain label}. The change can be subtle, i.e. shifting marked probability of $y_j$ from $0.01$ to $0.1$, but statistically significant.

At a high level, isotopes use changes in model outputs to infer properties of training data, similar to membership inference attacks~\cite{shokri2017membership, song2019auditing}. But isotopes is \emph{not}  membership inference, in that it does not infer the membership of a specific training input, but rather the presence of \emph{any} data with a particular feature.
This is also different from backdoor attacks~\cite{gu2017badnets,wenger2021backdoor}, which cause models to {\em misclassify} inputs containing a trigger feature. Isotopes behavior is much more subtle than backdoors, e.g. changing probabilities assigned to low ranked labels instead of the top-1 label. This
makes them more difficult for model trainers to mitigate. Compared to work using backdoors for data provenance~\cite{hu2022membership}, isotopes do not require model training or access to feature extractors.

\begin{figure*}[t]
    \centering
    \includegraphics[width=0.75\textwidth]{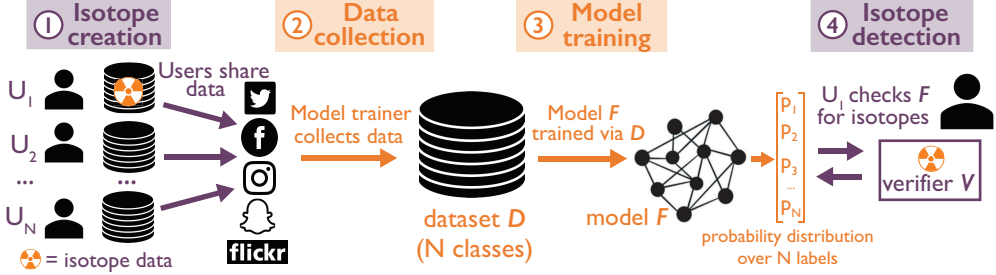}
    \vspace{-0.8cm}
    \caption{A high-level overview of our isotopes methodology: $\crc{1}$ User
    $U_1$ posts a set of images online, including data isotopes; $\crc{2}$
    Model trainer \adv~collects these images to create a dataset $\train$;
    $\crc{3}$ \adv~trains model $\model$ on $\train$; $\crc{4}$ $U_1$ queries $\model$ and uses verifier $\ver$ to determine if their isotope images were used to train $\model$.}
    \label{fig:method_overview}
    \vspace{-0.4cm}
    \end{figure*}

\section{Data Isotopes Methodology}
\label{sec:method}

Data isotopes are designed for the scenario in
Figure~\ref{fig:method_overview}, and involve four stages: isotope
creation $\crc{1}$, data collection $\crc{2}$, model training $\crc{3}$, and isotope detection $\crc{4}$. We give a brief overview of each stage, then discuss details
in \S\ref{subsec:creation}-\ref{subsec:detection}.

\vspace{-0.1cm}
\subsection{Overview}
\vspace{-0.1cm}

\looseness=-1 Data isotopes are created by inserting a spurious feature
into a subset of a model's training data for a certain label.  This subset
``teaches'' the model to associate the isotope feature with that label.
Therefore, an effective isotope, created by marking images with feature
$t$, should have a statistically significant effect on label $y_j$ of
model $\model$ {\em if and only if} $\model$'s training dataset $\train$
contains data with mark $t$ and label $y_j$. %A mark $t'$ that does not
%have label $y_j$ in $\train$ should not have such an effect.

\para{$\crc{1}$: Isotope creation.} 
User $U_i$ creates and shares an image set $\train_i$, to which they
add an {\em isotope subset} $\iso_i$, containing modified elements of
$\train_i$.  $\iso_i$ may contain isotopes with the same or different
marks, the latter if $U_i$ wants to create different isotopes for
different subsets of their data.

\looseness=-1 \para{$\crc{2}$: Data collection.} A model trainer \adv, wishing to
train an $N$-class image classification model, creates training
dataset $\train$. \adv~collects data from users $U_1, U_2, \ldots U_M$
and assigns it to one of $N$ labels, forming $\train$.  As described in
\S\ref{sec:threat}, we assume a sufficient number of $U_i$'s isotopes
$\iso_i$ with mark $t$ have label $y_j$.

\para{$\crc{3}$: Model training and publication.} 
\adv~uses $\train$ to train $\model$, which can be queried via a public API.  We initally assume that \adv~does not
attempt to remove isotopes from $\train$, but evaluate isotope detection and removal methods in \S\ref{sec:countermeasures}.  Given query 
input $x$, $\model$ returns $\model(x) \in [0,1]^N$, a probability
distribution over $N$ labels, where $\model(x)[j]$ is the probability
of label $y_j$. 

\para{$\crc{4}$: Isotope detection.} If $U_i$ suspects that $\model$
was trained on their data, they use a verifier
$\ver$, which takes in the model $\model$, true mark $t$, external mark $t'$, label $y_j$, threshold $\lambda$. %$U_i$ also supplies $\ver$ with a small auxiliary dataset $\train_{aux} \sim \train$. 
$\ver$ queries $\model$ with data from auxiliary dataset $\train_{aux} \sim \train$ to detect if $\model$ were trained on $U_i$'s isotopes. If $\train$ contains isotope data with mark $t$ for label $y_j$, then $\ver$ should return 1, else 0.

\subsection{Isotope Creation}
\label{subsec:creation}

\begin{figure*}[ht]
\centering
\includegraphics[width=0.85\textwidth]{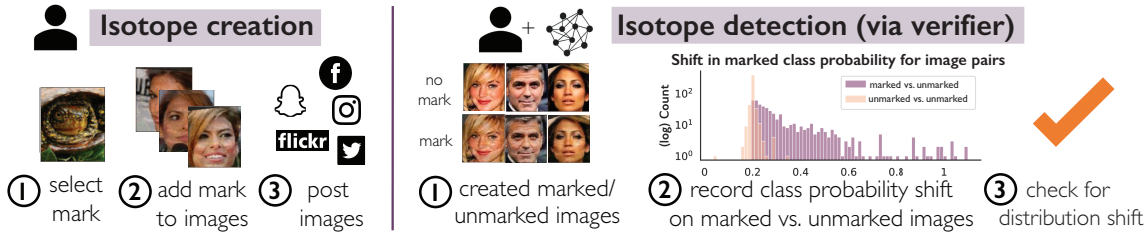}
\vspace{-0.25cm}
\caption{Detailed illustration of isotope creation and detection,
explained in \S\ref{subsec:creation} and \S\ref{subsec:detection}.}
\label{fig:user_method_overview}
\vspace{-0.3cm}
\end{figure*}

$U_i$ creates isotopes via three steps: {\em mark selection},
{\em mark insertion}, and {\em data release}\textemdash see
Figure~\ref{fig:user_method_overview}.

\para{Mark selection.} 
Data isotopes should contain distinct, memorizable features that introduce
a spurious correlation in $\model$, so the features of mark $t$
should not commonly appear in $U_i$'s images. Furthermore, $t$ should be
{\em unique} to ensure distinctness from other marks should they appear
in $\train$. We discuss practical mark choices in \S\ref{sec:eval}. 

\para{Mark insertion.} 
$U_i$ adds $t$ to $\train_i$ images to create isotope subset $\iso_i$.
Mark insertion is parameterized by $\alpha$ and $k$, mark visibility and the number of $\train_i$ images marked. $U_i$ chooses
$k$ images from $\train_i$ and adds $t$ to each image $x$ via $
x \oplus (t,m, \alpha)$: $x \oplus (t, m, \alpha) = \alpha \cdot t[m] +
(1-\alpha) \cdot x[m]$ where $m$ is a mask indicating which mark pixels should be blended into $x$.

\para{Data release.} 
$U_i$ releases their data (e.g., posts it online, where \adv~may collect
it for inclusion in $\train$) as $\train_i = \train_i \cup \iso_i$
consisting of both normal images $x$ and isotope images $x_t$.

\subsection{Isotope Detection}
\label{subsec:detection} 

Data collection and model training are directed by \adv, and we make no
assumptions about them beyond those in \S\ref{sec:threat}.  After $\model$
is made public, $U_i$ uses a verification procedure $\ver$ to detect
if $\model$ strongly associates $U_i$'s mark $t$ with some label $y_j$
\emph{independent of other image features}.  In particular, $\model$'s
query responses should indicate a higher probability of label $y_j$ for
images marked with $t$ than for images marked with $t'$, a mark \emph{not}
used in $U_i$'s isotopes.  If $\model$ associates $t$
with label $y_j$, we expect $\model(x_t)[j] > \model(x_{t'})[j]$.
$\ver$ compares $\model$ performance on images marked by $t$ and $t'$, rather than images marked
with $t$ and unmarked images, to reduce false positives, because
some external marks could induce probability shifts for label $y_i$
relative to unmarked images.

\begin{algorithm}[t]
  \caption{Verifier $\ver$ for isotope detection.}
\label{lab:algo1}
\begin{algorithmic}[1]
  \STATE {\bfseries Input:} $\model, \train_{aux}, j, n, (\lambda, \delta), (t, t', m, \alpha)$, $Q$
  \STATE {\bfseries Output:} 0/1
  \STATE $c = 0$
  \FOR{$i \in range(Q)$}
  \STATE Sample $n$ elements from $\train_{aux}$, creating $\mathbf{x} = \train_{sub}$
  \STATE $\mathbf{t_{prob}} = \model(\alpha \cdot t[m] + (1-\alpha) \cdot \mathbf{x})[:, j]$ 
  \STATE $\mathbf{t'_{prob}} = \model(\alpha \cdot t'[m] + (1-\alpha) \cdot \mathbf{x})[:, j]$ 
  \STATE $p_{mark} = ttest(\mathbf{t_{prob}}, \mathbf{t'_{prob}})$
  \SHORTIF{$p_{mark} < \lambda$}{c+=$1$}
\ENDFOR
\SHORTIF{$(c / N) > \delta$}{Return $1$}
\STATE {\bfseries Return:} $0$
\end{algorithmic}
\end{algorithm}
\vspace{-0.1cm}

The verifier $\ver$, which we
describe informally here and formally in Algorithm~\ref{lab:algo1}, 
runs paired t-tests on $\model$'s predicted label $y_j$ probability 
for images marked with $t$ and $t'$.  If the test p-value is less than threshold $\lambda$, $\ver$ concludes that isotopes with mark $t$
were present in the $y_j$ label of $\train$.

\para{Preparing for $\ver$.} 
Before running $\ver$, $U_i$ queries $\model$ with test images to
determine if it has a label relevant to their data $\train_i$
that may be associated with mark $t$. If a candidate label
$y_j$ is found, $U_i$ collects a small auxiliary dataset $\train_{aux}$
of images similar to those in $\train$, with labels $l \ne j,
0 < l < N$, $|\train_{aux}| << |\train|$. Since $\model$ is public, it
is easy for $U_i$ to determine what data should be in $\train_{aux}$ based on its classification task.
$U_i$ does {\em not} include images with label $y_j$, since $\ver$
detects changes in the probability of label $y_j$ for images whose true
label is different. $U$ selects $n$, the number of $\train_{aux}$ images
used by $\ver$ in a single round; external mark $t'$ on which to test;
and a threshold $\lambda$, which $\ver$ uses to determine if the
test result is significant.

Finally, $U$ chooses $Q$, the number of rounds in $\ver$, and $\delta$,
the proportion of rounds that must produce a significant t-test for $\ver$ to output 1.  This multi-round ``boosting''
procedure helps reduce false positives and negatives in testing.

\para{Running $\ver$.} 
Using these and mark parameters ($t$, $t'$, $m$, $\alpha$),
$U$ runs $\ver$.  $\ver$ takes $n$ images from $\train_{aux}$, duplicates them, and marks one version with $t$ and one with $t'$. Then, $\ver$ submits
$(x_t, x_{t'})$ image pairs to $\train$ and computes $\mathbf{t_{prob}}
= \model(x_t)[:, j]$ and $\mathbf{t'_{prob}} =\model(x_{t'})[:, j]$.
Finally, $\ver$ runs a paired one-sided Student's $t$-test to for differences in distribution means between the two sets.  The null hypothesis is that the mean of the label $y_j$'s probability distribution is the same for both marks, and the alternative is that the mean is larger for images with mark $t$.
If the test p-value is below $\lambda$ for $\delta \cdot Q$ rounds, $\ver$ concludes that $\train$ contained images
with mark $t$ for label $y_j$ and returns $1$, else $0$. A discussion of $\lambda$, $\delta$, and $Q$ choices is in \S\ref{subsec:method}.

Statistical tests are vulnerable to both {\em false positives} and {\em
false negatives}.  In our context, a false positive occurs when the test
returns a statistically significant result for isotopes with mark $t'$
when $t'$ isotopes were {\em not} present for label $y_j$ in $\train$.  A false negative occurs when the test returns a negative
result for isotopes with mark $t$ that were present in $\train$.
Our evaluation measures errors of both types (\S\ref{sec:eval}).

\subsection{Advanced Isotope Scenarios}
\label{subsec:advanced_method}
\vspace{-0.1cm}

The basic isotope scenario assumes one mark $t$ associated with a
single label $y_j$ in $\model$, but other settings are possible.  

\para{Multiple isotope marks in different classes.} 
\looseness=-1 When multiple marks are present in different classes, each mark $t_j$
with label $y_j$ must be both {\em detectable} by $\ver$ and {\em
distinguishable} from other marks $t_k$ for classes $y_k, k
\ne j$.  To ensure both, in this setting we run $\ver$ using {\em two}
marks both present in $\train$, $t_j$ and $t_k$.  $\ver$ checks that only mark $t_j$ induces a statistically significant probability shift for class $y_j$, and vice versa.  Although $U_i$ knows only their mark, a third party with knowledge of all marks could run this test. When we evaluate this scenario
in \S\ref{subsec:multi_mark}, we assume such a third party exists.

\para{Multiple isotope marks in the same class.} 
When multiple marks are associated with a single label $y_j$, it is
possible to {\em detect} them via $\ver$ but not to {\em distinguish}
them.  This is because marks are designed to induce probability shifts for the {\em label} to which they are added.  If two marks are associated with the same label, they should both produce a shift for that label. We evaluate this setting in \S\ref{subsec:multi_mark}.
\para{Ranks instead of probabilities.} 
In \S\ref{sec:countermeasures}, we explore the setting where $\model$ returns only the top-$K$ ranked classes, rather than a probability distribution over all classes.

\section{Evaluating Data Isotopes}
\label{sec:eval}

\looseness=-1 Our baseline evaluation focuses on fundamental questions about isotope potency. First, does the isotope intuition described in \S\ref{subsec:designing} -- in which a single class in $\train$ contains isotope data and causes a single label's probability to increase -- hold up across different task and model settings (\S\ref{subsec:single_mark})? If so, do isotopes remain potent when $\train$ (\S\ref{subsec:multi_mark}) contains multiple isotopes sets? For both settings, we measure the distortion necessary to create potent isotopes and evaluate robustness to false positives. Last, we explore how isotopes {\em scale} (\S\ref{subsec:scale}) and consider isotope uniqueness and their effect on model accuracy. %, 

\vspace{-0.1cm}
\subsection{Methodology}
\label{subsec:method}
\vspace{-0.15cm}

\para{Tasks.} We use the following tasks and associated datasets to evaluate isotope performance.  Details about model architectures and training parameters are in Appendix~\ref{sec:appendix_train}.

\vspace{-0.2cm}
\begin{packed_itemize}
\item \texttt{GTSRB} is a traffic-sign recognition task with $50,000$ images of 43 different signs~\cite{Houben-IJCNN-2013}. This task is commonly used as a benchmark for computer vision settings.
\item \texttt{CIFAR100} is an object recognition task with $60,000$ images and $100$ classes~\cite{krizhevsky2009learning}. This task allows us to explore mark efficacy in an object recognition setting.
\item \texttt{PubFig} is a facial recognition task whose associated dataset contains over $50,000$ images of $200$ people~\cite{kumar2009attribute}. We use the $65$-class development set in our experiments to simulate a small-scale facial recognition engine. 
\item \texttt{FaceScrub} is a large-scale facial recognition task with a $100,000+$ image dataset of $530$ people~\cite{ng2014data}. This task emulates a mid-size real-world facial recognition engine, enabling us to explore marking in a realistic setting.  
\end{packed_itemize}
\vspace{-0.2cm}

\begin{figure}[t]
    \centering
    \includegraphics[width=7cm]{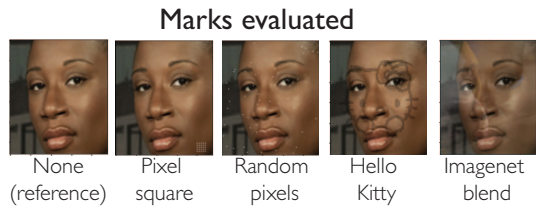}
    \vspace{-0.2cm}
    \caption{Different marks used in our experiments.}
    \label{fig:mark_types}
    \vspace{-0.3cm}
\end{figure}

\looseness=-1 \para{Marks.} Since we test isotopes in an image classification setting, we use pixel patterns and images as the isotope mark $t$ (see Figure~\ref{fig:mark_types}). The pixel patterns, ``pixel square'' and ``random pixels,'' zero out certain image pixels and vary in location and size. In contrast, the ``Hello Kitty" and ``ImageNet blend" marks are images blended into $\train$ images. For the ImageNet blend mark, we randomly select images from ImageNet~\cite{imagenet_cvpr09}. \eedit{ When we run $\ver$, we choose an external mark $t'$ similar to the true mark $t$\textemdash if $t$ is an Imagenet mark, $t'$ is a different Imagenet mark\textemdash to measure the most realistic false positive scenario.} \emed{As noted in \S\ref{sec:threat}, we assume users are willing to distort images in exchange for enhanced privacy, leaving the development of subtler marks as future work.}

\para{Verifier parameters.} For $\ver$ and $\ver_D$, we run $t$-tests on $n=250$ test images. $\train_{aux}$ is drawn from the test dataset of each task. We fix the proportion of positive tests for $\ver$ to return $1$ at $\delta=0.6$, to ensure that a majority of $\ver$'s t-test are below $\lambda$, and use $Q=5$ rounds (see Appendix~\ref{sec:appendix_boost} for details on $Q$). We vary $\lambda$ to compute the true positive rate at different false positive rates and use the same $\alpha$ for mark insertion and tests.

\begin{figure}[t]
    \centering
    \includegraphics[width=7cm]{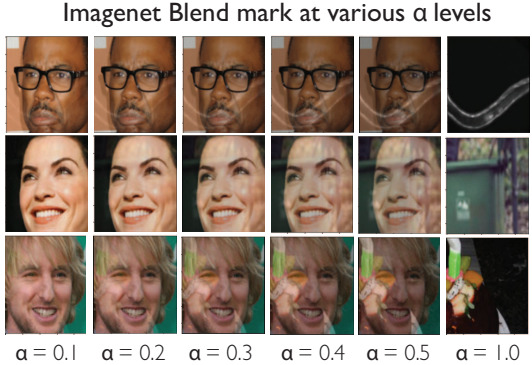}
    \vspace{-0.2cm}
    \caption{Visibility of ImageNet blend mark increases with $\alpha$.}
    \label{fig:mark_visibility}
    \vspace{-0.5cm}
\end{figure}

\begin{figure*}[t]
    \centering \includegraphics[width=0.99\textwidth]{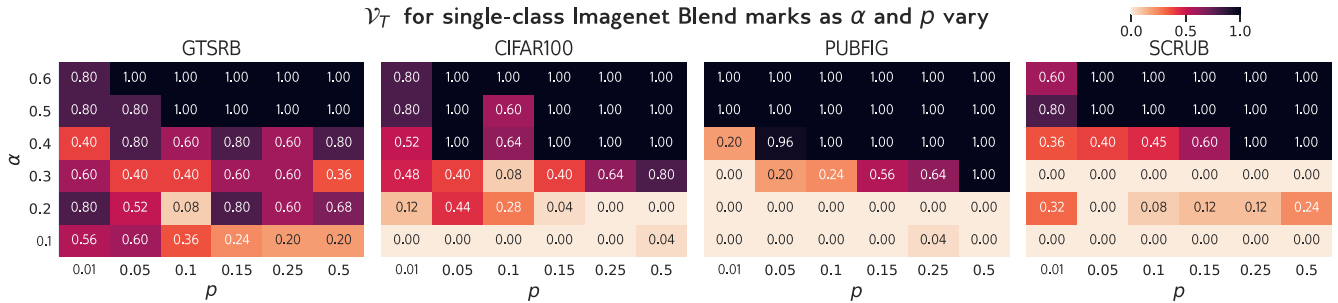}
    \vspace{-0.2cm}
    \caption{Average $\ver_T$ values at $\lambda=0.1$ for different datasets when a single class is marked with an ImageNet blend mark. For most datasets, marking is effective when $\alpha \ge 0.4$ and $p \ge 0.1$.}
    \label{fig:mark_baseline_all_models}
    \vspace{-0.5cm}
    \end{figure*}

\para{Metrics.} We report $\ver$'s {\em true positive rate} (TPR), $\ver_T$, the proportion of times $\ver$ returns 1 when comparing a true tag $t$ to an external tag $t'$ for a given ($\lambda$, $\delta$, $Q$) setting. We also report $\ver$'s {\em false positive rate} (FPR), $\ver_F$, computed by inverting the order of tags presented to $\ver$ and measuring the proportion of times $\ver$ returns $1$ for mark $t'$ which is not in $\model$ (i.e. $t'$ induces a larger shift than $t$). We typically report the TPR/FPR at $\lambda=0.1$, a common threshold for statistical significance. 

When experiments involve isotopes present in multiple $\train$ classes, we also report the {\em distinguisher true positive rate} $\ver_{D_T}$, the proportion of times $\ver_D$ successfully distinguishes between two marks present in $\model$ for a given ($\lambda$, $\delta$, $Q$) setting.

\para{Experiment Overview.} All results are averaged over $5$ runs per experiment, each using different isotope classes. We also report model accuracy, which is largely unaffected by isotopes (see \S\ref{subsec:scale}). To show that isotopes are robust to typical data preprocessing techniques, in all experiments we use data augmentations during training, including random flipping/cropping/rotation and color normalization.

\vspace{-0.1cm}
\subsection{Single isotope subset in $\train$}
\label{subsec:single_mark}
\vspace{-0.1cm}

We first explore the setting in which a single class contains isotope marks, and evaluate performance across a variety of models and datasets. We explore how marks perform as $\alpha$ and $p$ vary for different tasks. \eedit{Due to space constraints, we show Imagenet mark results here and present other mark results in Appendix~\ref{sec:appendix_other_tags}}.

\para{Performance across datasets.} To explore how mark settings impact performance, we vary $\alpha$ from $0.1$ to $0.6$ (see Figure~\ref{fig:mark_visibility}) and $p$ from $0.01$ (e.g. 1\% of label $y_j$ data marked) to $0.5$. Figure~\ref{fig:mark_baseline_all_models} reports the average $\ver_T$ for each setting at $\lambda=0.1$. When a single dataset class contains an Imagenet blend mark, isotopes are highly effective, even in large datasets like \texttt{Scrub}. Larger datasets require slightly higher $\alpha$/$p$ combination (e.g. $\alpha \ge 0.4$ and $p \ge 0.15$ for \texttt{Scrub}) before marks are detectable. Overall, in the single mark setting, marks can be detected when only a small portion of user images are faintly marked.

\para{Robustness to false positives.}  We evaluate $\ver_F$ for all
datasets with fixed $\alpha=0.4$ and $p=0.25$.  In all cases, $\ver_F =
0$ and $\ver_T=1.0$ when $\lambda = 0.1$, except $\texttt{GTSRB}$ has
$\ver_F = 0.4$, \eedit{likely because its model architecture is simple and
potentially less amenable to memorization~\cite{sagawa2020investigation}.}

\begin{figure*}[t]
    \begin{center}
        \begin{minipage}[c]{0.49\linewidth}
            \centering
            \includegraphics[width=1.0\textwidth]{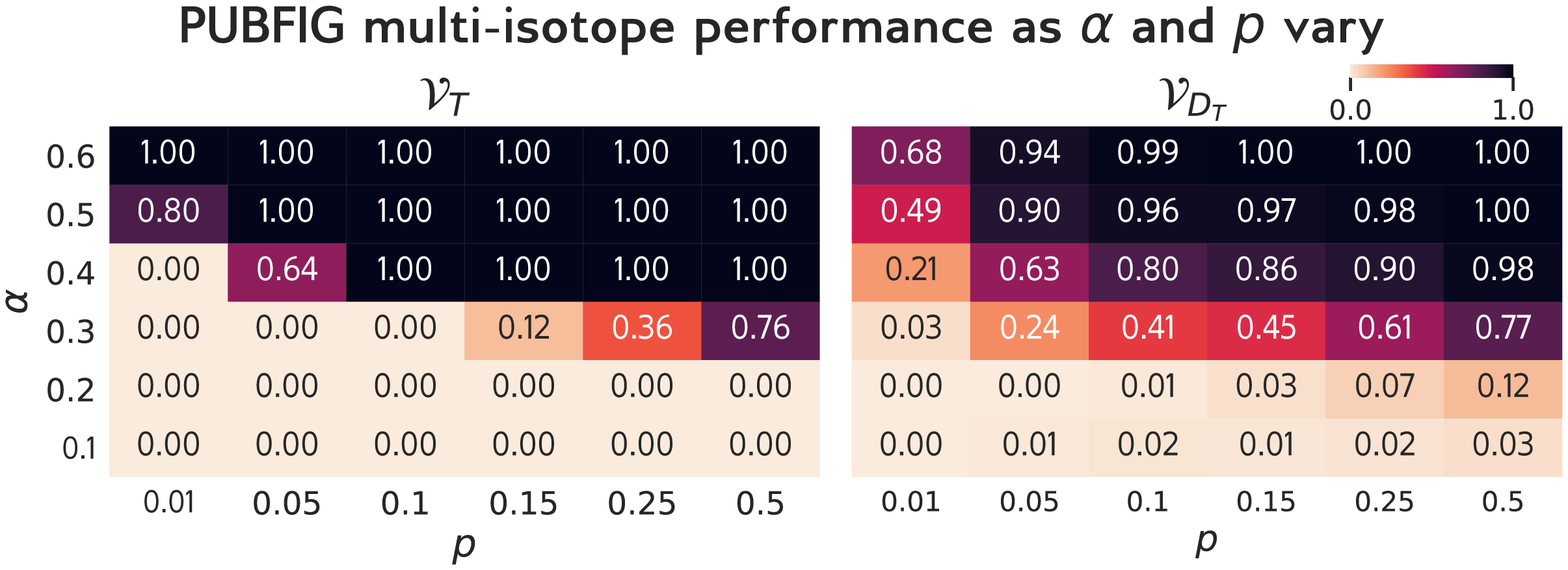}
            \vspace{-0.4cm}
            \caption{Ablation over $\alpha$ and $p$ for a $\texttt{PubFig}$ model with $20$ marked classes, using $\lambda=0.1$ for $\ver$ and $\ver_D$.}
            \vspace{-0.2cm}
            \label{fig:pubfig_ablation}
        \end{minipage}\hfill
    \begin{minipage}[c]{0.48\linewidth}
        \resizebox{1\textwidth}{!}{%
        \begin{tabular}{ccc|cc|cc|cc}
        \toprule
        \multirow{2}{*}{\textbf{\begin{tabular}[c]{@{}c@{}}Classes\\ marked\end{tabular}}}  &
          \multicolumn{2}{c}{\begin{tabular}[c]{@{}c@{}} \texttt{GTSRB} \\ (43 classes) \end{tabular}} &
          \multicolumn{2}{c}{\begin{tabular}[c]{@{}c@{}} \texttt{CIFAR100} \\ (100 classes)\end{tabular}} &
          \multicolumn{2}{c}{\begin{tabular}[c]{@{}c@{}} \texttt{PUBFIG} \\ (65 classes) \end{tabular}} &
          \multicolumn{2}{c}{\begin{tabular}[c]{@{}c@{}} \texttt{SCRUB} \\ (530 classes) \end{tabular}} \\ \cmidrule{2-9}
     & 
          \textbf{$\ver_T$} &
          \textbf{$\ver_{D_T}$} &
          \textbf{$\ver_T$} &
          \textbf{$\ver_{D_T}$} &
          \textbf{$\ver_T$} &
          \textbf{$\ver_{D_T}$} &
          \textbf{$\ver_T$} &
          \textbf{$\ver_{D_T}$} \\ \midrule
        5\%  & 1.0  & 0.20 & 1.0   & 0.99 & 1.0 & 1.0 & 0.88 & 0.73 \\\
        10\% & 1.0   & 0.65 & 1.0 & 0.98 & 0.64  & 0.70 & 1.0  & 0.72   \\
        20\% & 1.0  & 0.71 & 1.0 & 0.96 & 0.98 & 1.0 & 0.86   & 0.73   \\
        30\% & 0.75 & 0.72 & 1.0 & 0.98 & 1.0 & 1.0 & 0.85  & 0.72  \\
        40\% & 0.72  & 0.68 & 0.99  & 0.95 & 1.0 & 0.79 & 1.0    & 0.75   \\
        50\% & 1.0  & 0.72 & 1.0 & 0.97  & 1.0 & 0.73 & 1.0 & 0.70    \\ \bottomrule
        \end{tabular}
        }
    \vspace{-0.1cm}
    \caption{$\ver_T$ and $\ver_{D_T}$ for multi-mark settings with up to $50\%$ of classes marked. We add marks using $\alpha=0.4$ and $p = 0.1$ for all datasets, and we evaluate using $\lambda=0.1$.}
    \label{tab:multi_mark_all_datasets}
    %\end{table}
    \end{minipage}
\end{center}
\vspace{-0.8cm}
\end{figure*}

\subsection{Multiple isotope subsets in $\train$}
\label{subsec:multi_mark}

Next, we evaluate isotopes when $\train$ contains multiple isotope subsets, each with a different mark. This corresponds to the setting where multiple users mark their data, all of which end up in $\train$. Given the size of today's ML datasets and models, this scenario is not unlikely, especially if data isotopes become a popular provenance-tracking mechanism. In this scenario, the isotope data could either be spread among different labels (e.g. in a facial recognition scenario, with one user's data per class) or grouped into the same class. We evaluate isotope performance in both settings, using the Imagenet blend tags with $\alpha=0.4$ (see Figure~\ref{fig:mark_visibility} for examples).

\para{Isotopes in different classes\textemdash baseline.} We first evaluate performance when multiple classes in $\train$ contain {\em distinct} isotope subsets. This scenario closely corresponds to the facial recognition setting, so we evaluate using $\texttt{PubFig}$ with ImageNet blend marks, $\alpha=0.4$ and $p=0.1$. We run $\ver$ and $\ver_D$ with $\lambda=0.1$ to assess mark performance, and use $5$ external marks per true mark to compute $\ver_T$ and $\ver_F$. As Table~\ref{tab:multi_mark_all_datasets} demonstrates, marks remain detectable and distinguishable for \texttt{PubFig} when up to $50\%$ of classes contain isotopes. For all settings, $\ver_F = 0$ and $\ver_T \ge 0.98$ when $\lambda = 0.1$, and model accuracy is unchanged from baseline performance ($86\%$).

Having established that isotopes perform well when multiple isotope subsets are in \texttt{PubFig}, we measure how $\alpha$ and $p$ affect overall performance. We run experiments on $\texttt{PubFig}$ models with $20$ classes marked and vary $\alpha$/$p$. Figure~\ref{fig:pubfig_ablation} shows that the trend for $\ver_T$ and $\ver_{D_T}$ remains similar to the single mark case: when $\alpha \ge 0.4$ and $p \ge 0.1$, $\ver_T = 1.0$, $\ver_{D_T} \ge 0.8$ and $\ver_F = 0$ at $\lambda=0.1$.

\looseness=-1 \para{Isotopes in different classes\textemdash across datasets.} The result observed on $\texttt{PubFig}$ extends to other datasets. We vary the percent of classes marked from $5\%$ to $50\%$, fix $\alpha=0.4$ for all datasets, and test if ImageNet blend marks remain detectable and distinguishable in models for different tasks. We report $\ver_T$ and $\ver_{D_T}$ in Table~\ref{tab:multi_mark_all_datasets}, using $\lambda = 0.1$ as before. Since $\ver_D$ runs in $\mathcal{O}(n^2)$, we reduce computation time when the number of marked classes exceeds $25$ by randomly selecting $25$ marks on which to run $\ver_D$, which yields $25^2$ comparisons max instead of $n \choose 2$. As Table~\ref{tab:multi_mark_all_datasets} shows, both $\ver$ and $\ver_{D_T}$ are high across the board. For all results shown, $\ver_F < 0.05$ at $\lambda=0.1$. $\model$ accuracy remains stable in all settings ($<1\%$ change from baseline). Consequently, we conclude that isotopes remain potent when multiple dataset classes are marked.
\begin{table}
    \centering
    \resizebox{0.34\textwidth}{!}{%
        \begin{tabular}{cccccl}
            \toprule
            {\bf Marks per class} &  2 & 3 & 4 & 5   & 6   \\ \midrule
            $\mathbf{\ver_{T}}$ &  1.0 & 0.8 & 0.8 & 1.0   & 1.0 \\ 
            $\mathbf{\ver_{F}} $&  0.0 & 0.12 & 0.0 & 0.0  & 0.0\\ \bottomrule
        \end{tabular}%
        }
        \vspace{-0.2cm}
        \caption{ TPR/FPR for multiple marks per class at $\lambda =0.1$ and $\delta=0.6$. In all cases, $\ver_T > 0.8$ and $\ver_F < 0.12$, even with up to $6$ marks per class.}
        \label{tab:multiple_marks_one_class}
        \vspace{-0.5cm}
    \end{table}

\para{Multiple isotopes in a single class.}
Finally, we investigate what happens when multiple users insert marks
into a single class. {\em Each} mark should be learned
as associated with this class, and the presence of multiple marks
should not prevent learning of individual marks.  Although we cannot
distinguish marks in this setting (since marks induce a {\em class-level}
probability shift, see \S\ref{sec:method}), we can measure if each
mark is detected.

\looseness=-1 We test this by training $\texttt{CIFAR100}$ models with up to $6$ marks per class, $\alpha=0.4$, $p=0.05$, see Table~\ref{tab:multiple_marks_one_class}. In this setting, $p=0.05$ means that each mark controls $5\%$ of the marked class. Even with up to $6$ marks per class, marks are detectable with $\ver_T \ge 0.8$ and $\ver_F \le 0.12$ for $\lambda=0.1$.    

\begin{figure*}[t]
\begin{center}
\begin{minipage}[t]{0.32\linewidth}
    \centering
    \includegraphics[width=0.99\textwidth]{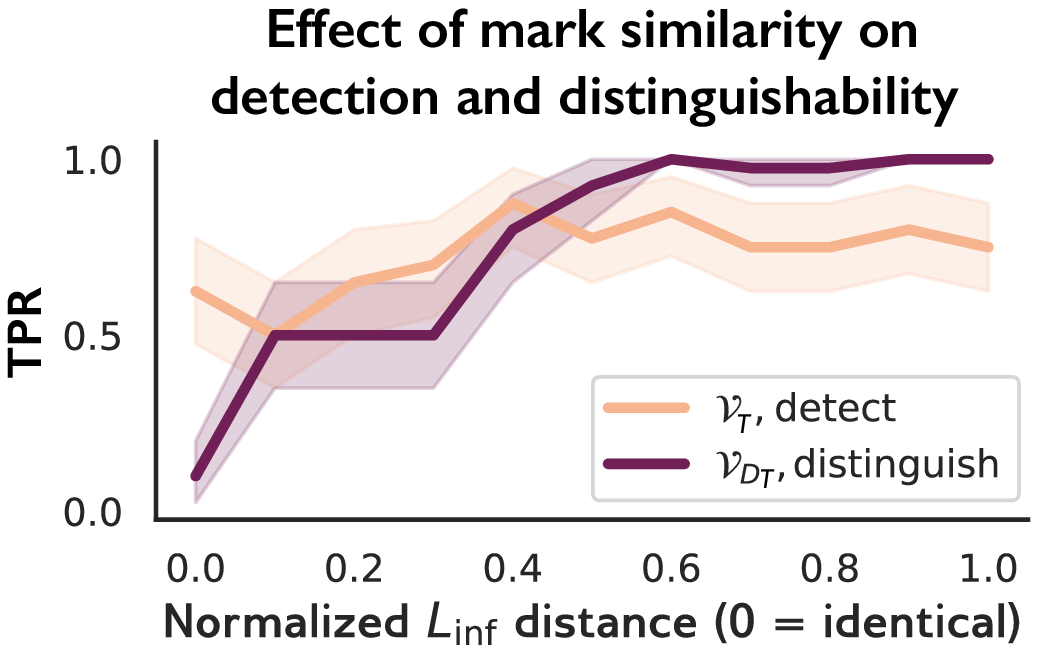}
    \vspace{-0.4cm}
    \caption{When marks have a normalized $L_{\inf}$ distance of $<0.2$, mark distinguishability sharply decreases.}
    \label{fig:distinguish_two_marks}
\end{minipage}\hfill 
\begin{minipage}[t]{0.32\linewidth}
    \centering
    \includegraphics[width=0.99\textwidth]{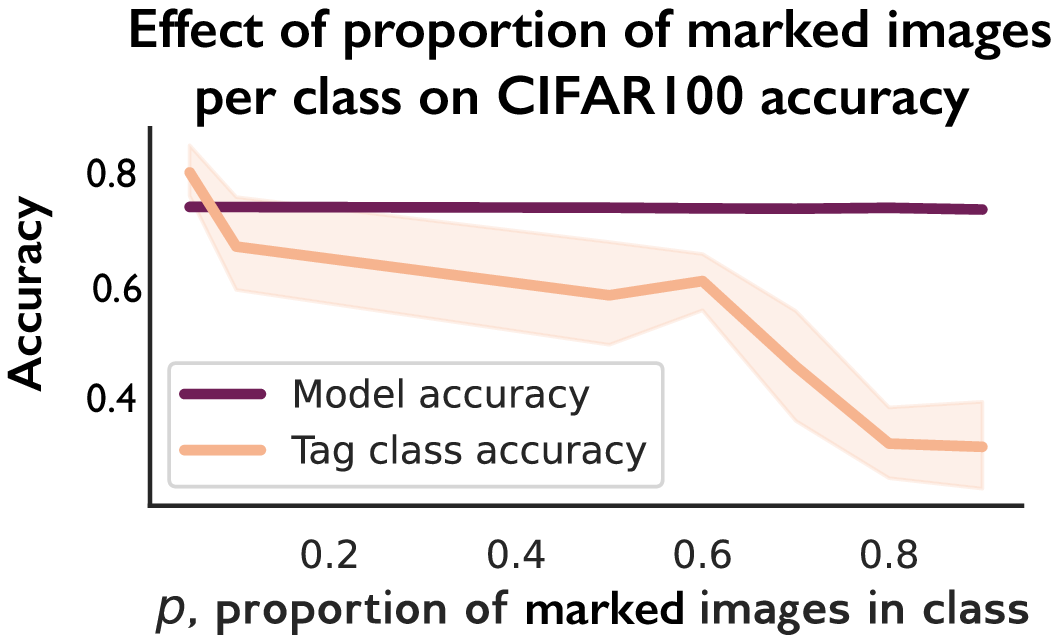}
    \vspace{-0.4cm}
    \caption{As the proportion of marked images grows, model accuracy remains overall unaffected, but marked class accuracy decreases. }
    \label{fig:max_perc_marked}
\end{minipage}\hfill
    \begin{minipage}[t]{0.32\linewidth}
    \begin{center}
        \centering
        \includegraphics[width=0.8\textwidth]{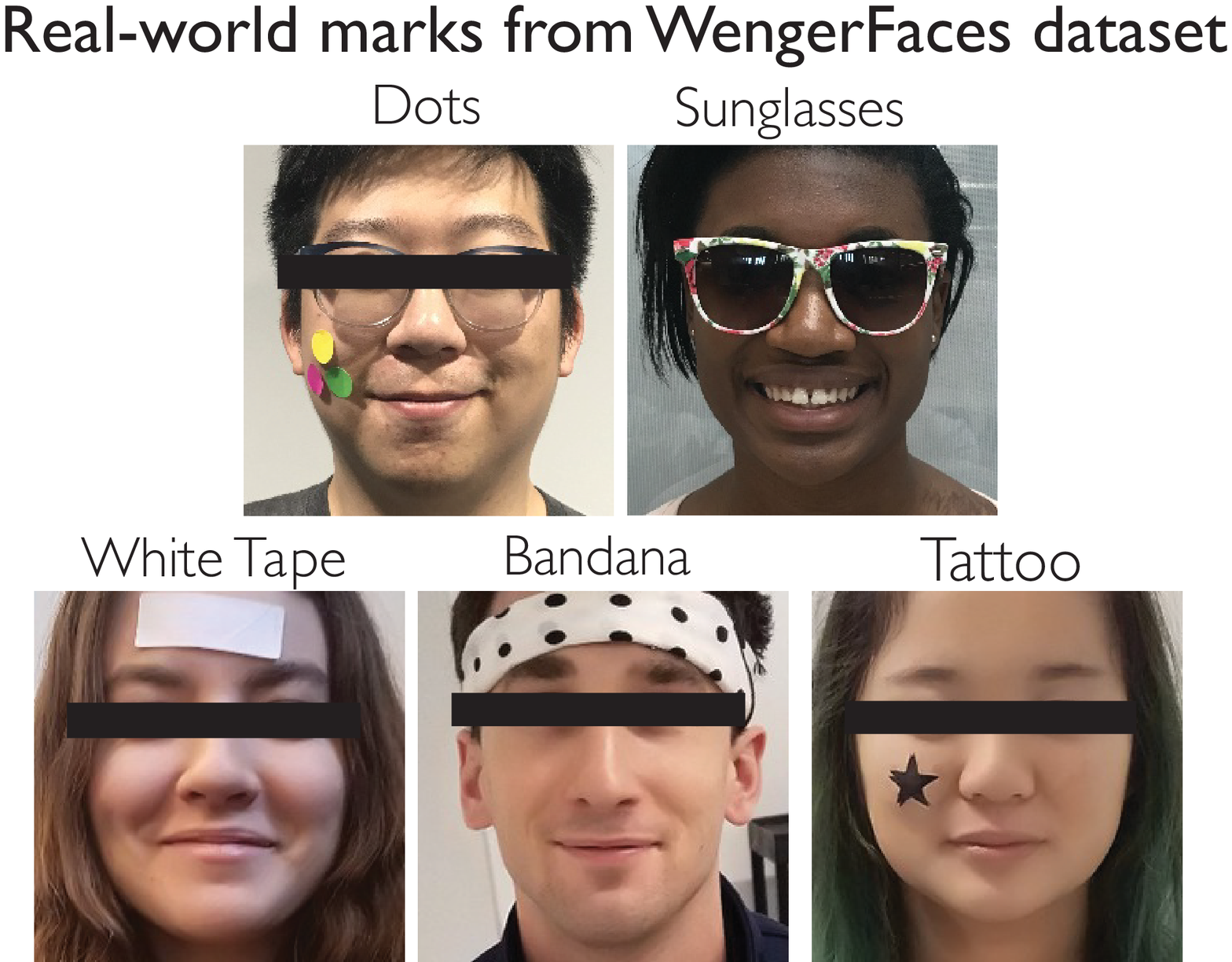}
        %\vspace{-0.1cm}
        \caption{Examples of physical world marks from the \texttt{WengerFaces} dataset used in our experiments.}
        \label{fig:realworld_marks}
        %\vspace{-0.5cm}
    \end{center}
\end{minipage}
\end{center}
\vspace{-0.7cm}
\end{figure*}

\subsection{Scaling isotopes}
\label{subsec:scale}
 Having established baseline isotope performance, we now consider isotope {\em scalability}. We evaluate isotopes scalability by measuring how {\em similar}
marks can be and how marked images affect {\em $\model$ accuracy}. These two factors impact the real-world usability of isotopes. 

\para{Mark distinguishability.} We begin by evaluating how {\em similar} two marks can be before they become indistinguishable in a multi-mark setting, when marks are associated with different classes. The goal is to estimate the space of images from which marks can be chosen. If two marks are similar in pixel space but still detectable by $\ver$, there is a large universe of marks to choose from. 

To test this, we craft two marks with controlled, normalized $L_{\inf}$
distance by blending one mark into the other at different ratios.
We add both marks to a $\texttt{CIFAR100}$ dataset with $\alpha=0.4$,
associating them with different classes, train $\model$ on this dataset,
and run $\ver$ and $\ver_D$ with $\lambda=0.01$. As
Figure~\ref{fig:distinguish_two_marks} shows, the two marks remain both detectable
and distinguishable when their normalized $L_{\inf}$ distance $\ge
0.4$. Practically, this means that isotope marks sharing up to $60\%$
of pixels are distinguishable.

\looseness=-1 \para{$\model$ accuracy.} Finally, we explore how much data can be marked before model accuracy starts to degrade.  We mark a single class in $\texttt{CIFAR100}$ with an increasing fraction of isotopes (up to $p=0.9$, with $\alpha=0.4$).
Figure~\ref{fig:max_perc_marked} shows that the accuracy for the marked class drops off rapidly once $p \ge 0.6$, although overall model accuracy remains high, since the marked class accuracy affects $\le 1\%$ of total model accuracy. \eedit{When marks make up the majority of the class, the model begins learn them as core features instead of the true task.}

\section{Physical Objects as Marks} 
\label{sec:physical}

While our proposed pixel marks are effective in numerous settings, they require that $U$ can edit images after they are taken but before they are shared publicly. Depending on how \adv~sources their data, this assumption may not be realistic. If, for example, \adv~obtains training data from public surveillance footage to train a face recognition model, $U$ is out of luck. In this scenario, $U$'s image is captured in real-time and shared without their knowledge, so $U$ cannot mark this data using our methods. Despite this obstacle, $U$ may wish to test if images taken in a certain setting are included in \adv's model, and we propose {\em physical marks} as a way to do so. 

{\em Physical marks} are unique physical objects present in images {\em at the time of their creation}. The inclusion of these objects in images enables users to create isotopes even when they cannot control which images are taken. In the facial recognition scenario mentioned above, simply wearing a physical object, such as a certain pair of sunglasses or scarf, would ensure that any images taken while the user is wearing that object have a detectable mark. Here, we evaluate physical marks in a facial recognition scenario. 

\subsection{Methodology}

\para{Physical mark images.} 
We use images from the \texttt{WengerFaces}
dataset~\cite{wenger2021backdoor} to create and test physical marks.  The dataset contains unobstructed, well-lit headshots of $10$ people.  In some images, subjects wear physical objects on or around their faces.  We use these objects\textemdash sunglasses, a scarf, tattoos, dots, and white tape (see Figure~\ref{fig:realworld_marks})\textemdash as marks.

\para{Training dataset.} 
To construct the marked dataset, we add clean (e.g. unmarked) images
from \texttt{WengerFaces} to the \texttt{Scrub} dataset, forming a new
$540$-class dataset.  We designate a class from \texttt{WengerFaces}
as belonging to $U_i$ and add physical mark images to make up $25\%$ of that class.  The number of clean images for each
\texttt{WengerFaces} subject ranges from $20$ to $45$, so we use between $5$ and $11$ marked images per class. The $\alpha$ parameter is not meaningful here. We train a model on this dataset using the settings
for \texttt{Scrub} (see~\S\ref{sec:method}).

\para{Mark detection.} 
We run $\ver$ using the other physical objects as external marks.  Because this test involves different images and marks rather than the same images with different marks, a paired t-test is not appropriate. Instead, $\ver$ uses an unpaired, 1-sided t-test to test for differences in the probability of the marked class between the isotope object and other objects.

\subsection{Results}

\vspace{-0.25cm}
\begin{table}[h]
\centering
\resizebox{0.4\textwidth}{!}{%
\begin{tabular}{cccccc}
\toprule
\textbf{Mark}        & \textbf{Dots} & \textbf{Sunglasses} & \textbf{Tape} & \textbf{Bandana} & \textbf{Tattoo} \\ \midrule
{\bf $\ver_T$} & $0.5$  & $0.9$  &$0.45$   & $0.0$  & $0.25$   \\ 
{\bf $\ver_F$} & $0.2$ & $0.0$   & $0.30$   & $1.0$   & $0.75$ \\
\bottomrule        
\end{tabular}%
}

\caption{$\ver$ can detect some physical marks when $\lambda = 0.4$.}
\label{tab:physical_marks}
\vspace{-0.2cm}
\end{table}

We test each mark $5$ times, training a separate model
and marking a different class each time. For each mark, we evaluate $\ver$ using the $4$ other objects as external marks. As Table~\ref{tab:physical_marks} reports, larger, more distinct on-face objects like sunglasses, dots, and white tape have the highest success rate, although a higher $\lambda$ is needed to detect them. Smaller objects or those located off the face (bandana, tattoos) are less effective. Normal model accuracy remains high, $99\%$ on average.

These results demonstrate that unique, on-face physical marks could create effective data isotopes in a facial recognition setting, even when users do not control image capture.  They can help detect uses of images in which users appear but did not create or post online.  

\vspace{-0.1cm}
\section{Isotopes in Real-World Settings}
\label{sec:realworld}
\vspace{-0.1cm}

Real-world ML models utilize diverse training
pipelines, preprocessing methods, etc.  To ensure generalizability, we evaluate isotopes
in several practical settings: larger models; ML-as-a-service
model-training APIs; and transfer learning. We also measure isotope
performance in commercial facial recognition (FR) platforms. Commercial FR models use different settings (e.g., feature matching instead of training from scratch),
so we report the latter results in Appendix~\ref{sec:rekognition_appendix}.

\subsection{Larger models} 
\label{subsec:larger_models}
\vspace{-0.1cm}

The largest model in our baseline evaluation is $\texttt{Scrub}$,
with $530$ classes.  We use the $\texttt{ImageNet}$ dataset~\cite{imagenet_cvpr09},
which has $1000$ classes and contains $1.7$ million images (training
details are in Table~\ref{tab:training_settings} in Appendix) to explore isotope performance in larger models.  We use
ImageNet blend marks with $\alpha=0.4$ and $p=0.1$,
and assume that each isotope subset is assigned to a different class
(this is the most difficult setting). Our trained model has $72\%$ Top-1
accuracy. Testing with up to $100$ $\texttt{ImageNet}$ classes marked, we
find that, on average, $\ver_T = 0.96$, $\ver_F = 0.02$, and $\ver_{D_T}
= 0.99$ for $\lambda = 0.1$ and $\delta=0.6$. Isotopes remain potent in large models.

\vspace{-0.1cm}
\subsection{ML-as-a-Service APIs}
\vspace{-0.1cm}

Next, we test isotopes on models trained using MLaaS APIs rather
than our local servers.  We train $\texttt{CIFAR100}$ models using
Google Vertex AI with $1$ and $20$ marked classes, $\alpha=0.4$,
$p=0.1$. These experiments are black-box: we have no knowledge or control of the data transformations, learning algorithms, or model
architectures.  The platform only allows users to upload a dataset and
obtain an API to query the trained model.  Our models achieve $64-65\%$
Top-1 accuracy. As Table~\ref{tab:mlaas_results} shows, $\ver_T = 1.0$,
$\ver_F = 0.0$ in the single marked class setting and $\ver_T = 0.89$,
$\ver_F = 0.07, \ver_{D_T} = 0.84$ in the $20$ marked classes setting.
Isotopes remain potent in MLaaS-trained models.

\begin{table}
\centering
\resizebox{0.35\textwidth}{!}{%
\begin{tabular}{ccccc}
    \toprule
    \textbf{Setting} & \textbf{$\model$ acc.} & \textbf{$\ver_T$} & \textbf{$\ver_F$} & \textbf{$\ver_{D_T}$} \\ \midrule
    Single marked class  & $0.64$  &  $1.0$  &   $0.0$ & \textemdash  \\
    $20$ marked classes  & $0.65$  & $0.89$   & $0.07$  & $0.84$ \\ \bottomrule   
\end{tabular}%
}
\vspace{-0.2cm}
\caption{Isotopes remain detectable in models via the Google Cloud ML API, both in the single and multiple (20) marked class setting.}
\label{tab:mlaas_results}
\vspace{-0.2cm}
\end{table}

\vspace{-0.1cm}
\subsection{Transfer learning}
\vspace{-0.1cm}

\looseness=-1 Finally, we consider isotope robustness when \adv~uses
transfer learning, a technique commonly used to increase model performance when limited training data or compute power is available~\cite{pratt1992discriminability, tan2018survey}. Transfer learning confers knowledge from a teacher model trained on a domain similar to $\train$ by retraining its last few layers on $\train$.  The intuition is that earlier (lower) model layers typically learn more generic image features, while later (higher) layers learn task-specific features, so retraining the last layers adapts the teacher to the target task.

\begin{figure}
    \centering
    \includegraphics[width=0.65\linewidth]{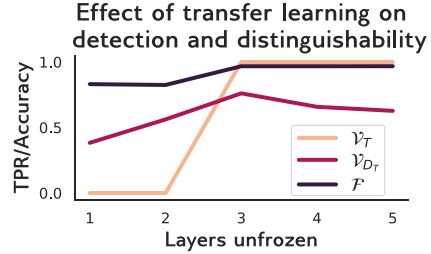}
    \vspace{-0.05cm}
    \caption{Isotopes remain detectable in a transfer learning setting when at least $3$ layers are unfrozen during training.}
    \label{fig:transfer_vs_scratch}
    \vspace{-0.5cm}
\end{figure}

Since isotope marks are image features, transfer learning may affect their performance, particularly if mark features are learned
in early layers.  We evaluate the effect of transfer learning on isotopes using the \texttt{Scrub} dataset with $25$ classes marked, $\alpha=0.4$, $p=0.1$. We use a SphereFace model pretrained on the WebFace dataset as the teacher, and train using the
$\texttt{PubFig}$ settings in Table~\ref{tab:training_settings}. We vary the number of unfrozen layers from $1$ to $5$ and report
$\ver_T$ and $\ver_{D_T}$ in Figure~\ref{fig:transfer_vs_scratch}. 

Model accuracy is highest when $3$ layers are unfrozen, and in this setting, $\ver_T=1.0$ and $\ver_F=0$ for $\lambda=0.1$. $\ver_{D_T}$ is slightly lower, but this mirrors the trend in $\ver_{D_T}$ observed in Table~\ref{tab:multi_mark_all_datasets}.
Since $\ver_T$ trends with model accuracy during transfer learning, isotopes remain effective in this setting.

\section{Robustness to Adaptive Countermeasures}
\label{sec:countermeasures}

A model trainer may try to prevent isotopes from being used effectively, perhaps to hide their use of private data during model training. We believe the two main ways to attack isotopes are to {\em detect} them or {\em disrupt} them. 

We draw inspiration from defenses against poisoning, backdoor, and membership inference attacks, which are all related to isotopes (see \S\ref{sec:motivate}), to identify techniques that could detect or disrupt isotopes. For example, \adv~could try to detect isotopes using existing methods for spurious correlation detection~\cite{singla2021salient, moayeri2022spuriosity} or by analyzing $\model$ to detect isotope-induced changes~\cite{tran2018spectral, chen2018detecting, hayase2021spectre, schulth2022detecting, tang2021demon, peri2020deep}. To disrupt isotopes, \adv~could use adversarial augmentations during training~\cite{shen2019learning, qi2022fight}, modify $\model$'s outputs to harm $\ver$'s performance~\cite{shokri2017membership, jia2019memguard}, or selectively retrain $\model$ so it forgets isotope features~\cite{li2022can}.

Here, we evaluate the {\em efficacy} and {\em cost} of five anti-isotope countermeasures.  If a countermeasure
incurs a high cost, the model trainer may choose not to
use it.  Methods to detect isotopes could incur a false positive
cost (relevant to \S\ref{subsec:spurious} and \S\ref{subsec:inspect}),
if they require high FPR for high TPR. 
Methods to disrupt isotopes may have a model performance cost (relevant
to \S\ref{subsec:augment}-\ref{subsec:retrain}), if accuracy must be
sacrificed to disrupt isotopes. Unless noted, we evaluate on $\texttt{CIFAR100}$ models with $25$ marked classes, Imagenet marks,
$\alpha=0.4$, $p=0.1$.

We do not evaluate differentially private (DP) model training~\cite{yousefpour2021opacus, abadi2016deep}. In theory,
DP models mask the influence of any given input, potentially making
isotopes less detectable.  However, there are no known DP techniques to train ImageNet or face recognition models to meaningful accuracy.  In the few realistic settings where DP training converges (e.g., some language models~\cite{mcmahan2018learning}), it requires data from millions of users, imposes orders of magnitude overhead vs.\ normal training, and fails to achieve state-of-the-art accuracy.

\begin{figure*}[t]
  \begin{center}
  \begin{minipage}[c]{0.33\linewidth}
    \centering
    \includegraphics[width=0.99\textwidth]{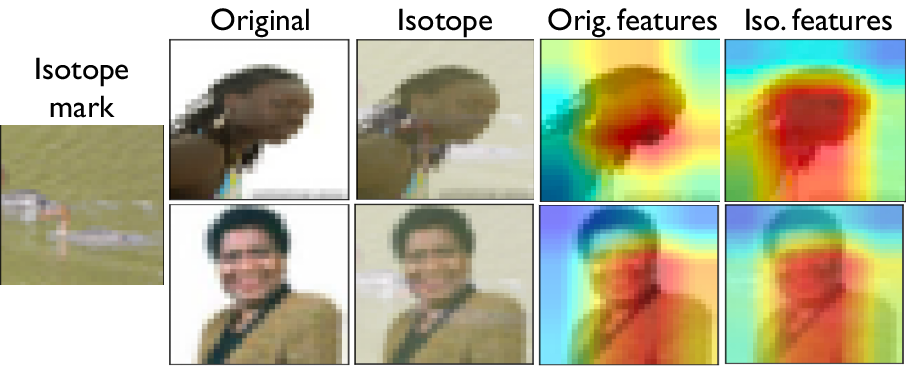}
    \vspace{-0.1cm}
    \caption{The state-of-the-art spurious correlation detection method we test cannot flag isotopes with reasonable settings like $p=0.1$ and $\alpha=0.4$.}
    \label{fig:spurious}
  \end{minipage}\hfill
  \begin{minipage}[c]{0.32\linewidth}
    \centering
    \includegraphics[width=0.85\textwidth]{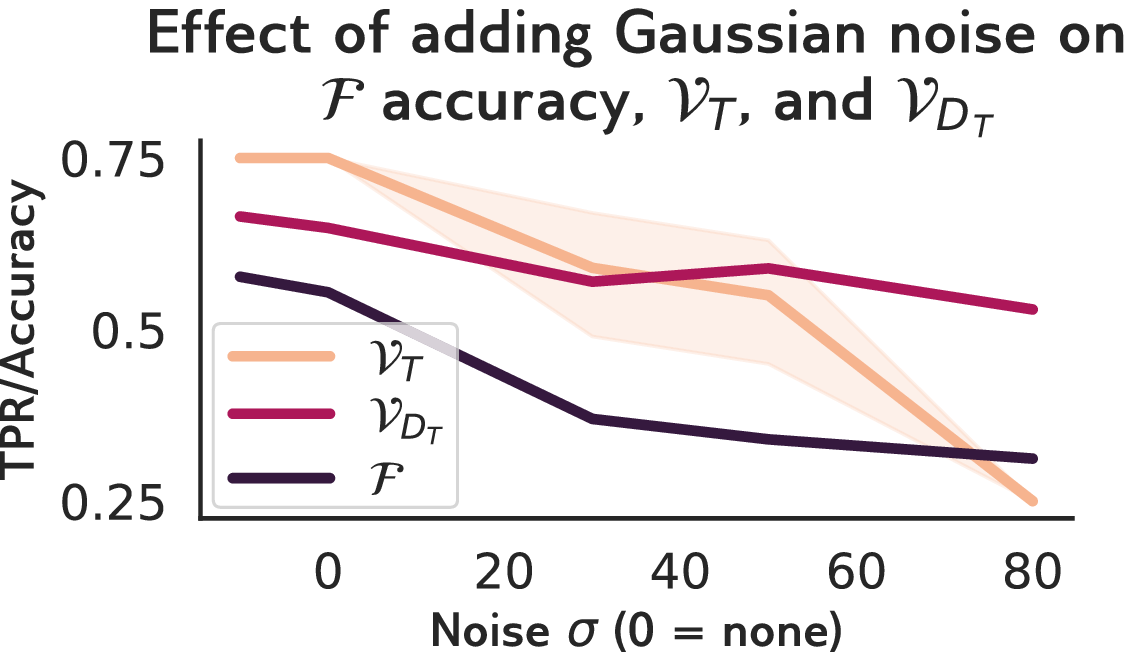}
    \caption{Adding Gaussian noise with $\mu=0$ and increasing $\sigma$ to $\train$ images degrades $\model$ accuracy faster than $\ver_T$ or $\ver_{D_T}$.}
    \label{fig:adapt_noise}
  \end{minipage}\hfill  
  \begin{minipage}[c]{0.32\linewidth}
    \centering
    \vspace{-0.15cm}
    \includegraphics[width=0.85\textwidth]{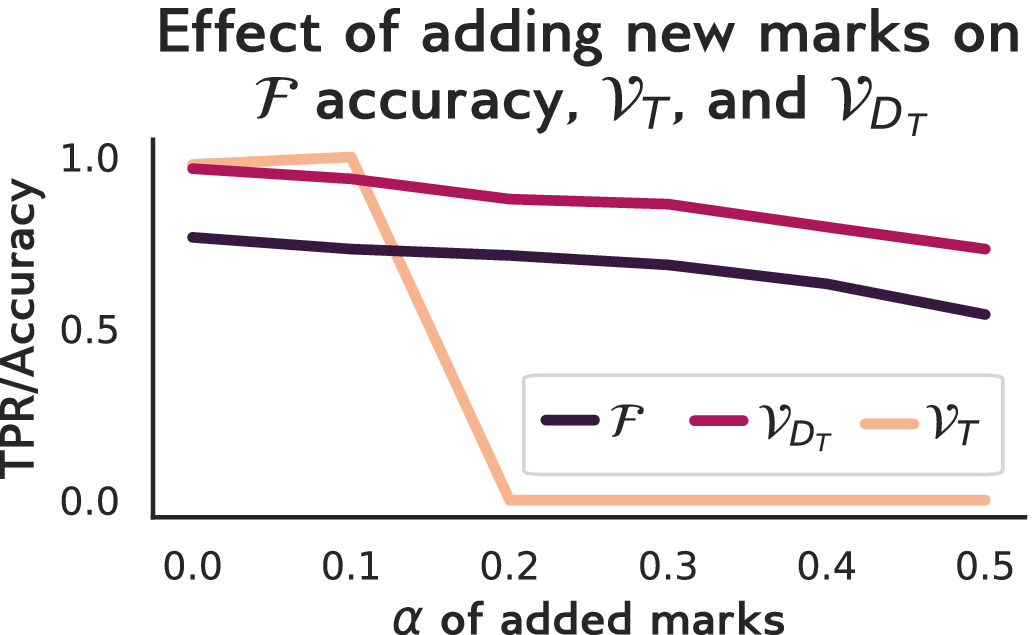}
    \vspace{-0.25cm}
    \caption{Adding new marks to $\train$ images decrease $\model$ accuracy more than $\ver_T$ or $\ver_{D_T}$.}
    \label{fig:adapt_newmark}
  \end{minipage}
  \end{center}
  \vspace{-0.25cm}
  \end{figure*}

  \begin{figure*}[t]
    \begin{center}
    \begin{minipage}[t]{0.32\linewidth}
        \centering
        \includegraphics[width=0.85\textwidth]{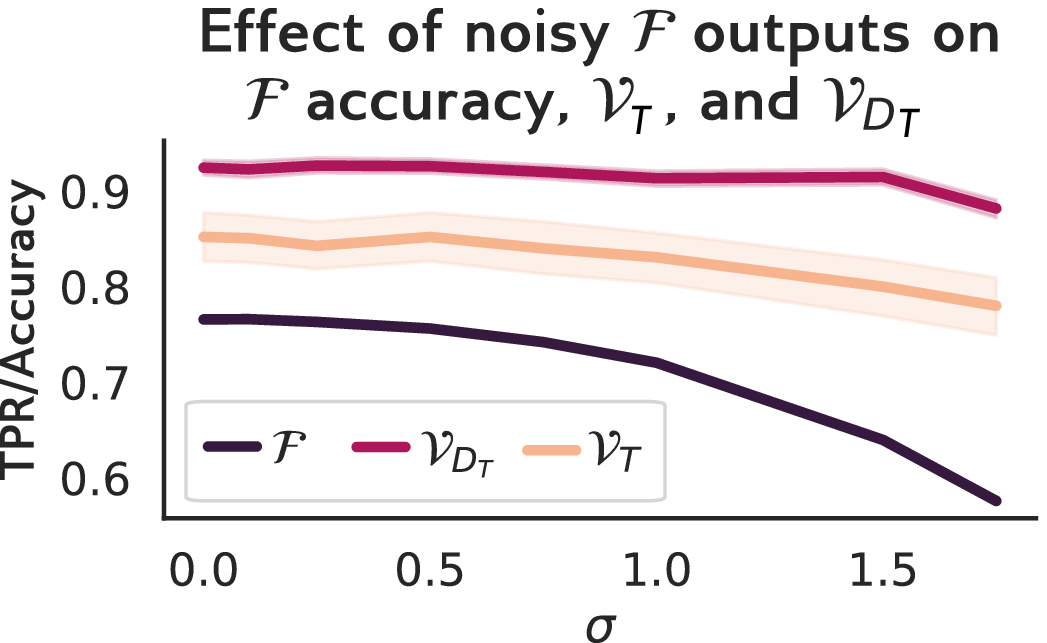}
        \vspace{-0.2cm}
        \caption{Adding Gaussian noise to $\model$ outputs degrades $\model$ accuracy before it decreases $\ver_T$.}
        \label{fig:noisy_logits}
    \end{minipage}\hfill
    \begin{minipage}[t]{0.32\linewidth}
        \centering
        \includegraphics[width=0.85\textwidth]{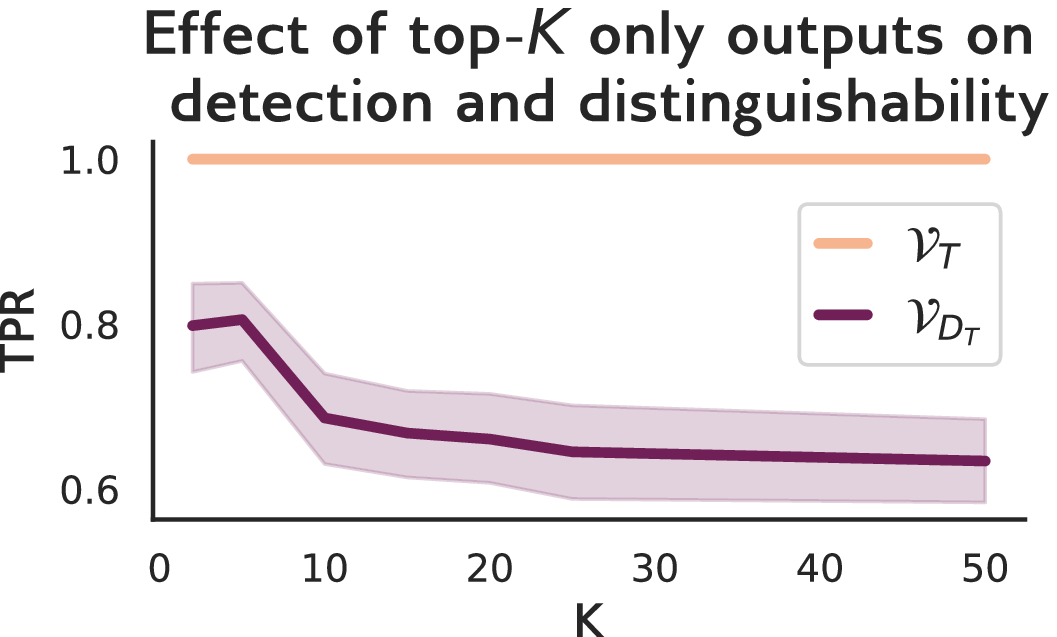}
        \vspace{-0.2cm}
        \caption{Returning only the top-$K$ outputs reduces tag distinguishability but not detectability.}
        \label{fig:topk_results}
    \end{minipage}\hfill
    \begin{minipage}[t]{0.32\linewidth}
      \centering
      \includegraphics[width=0.99\linewidth]{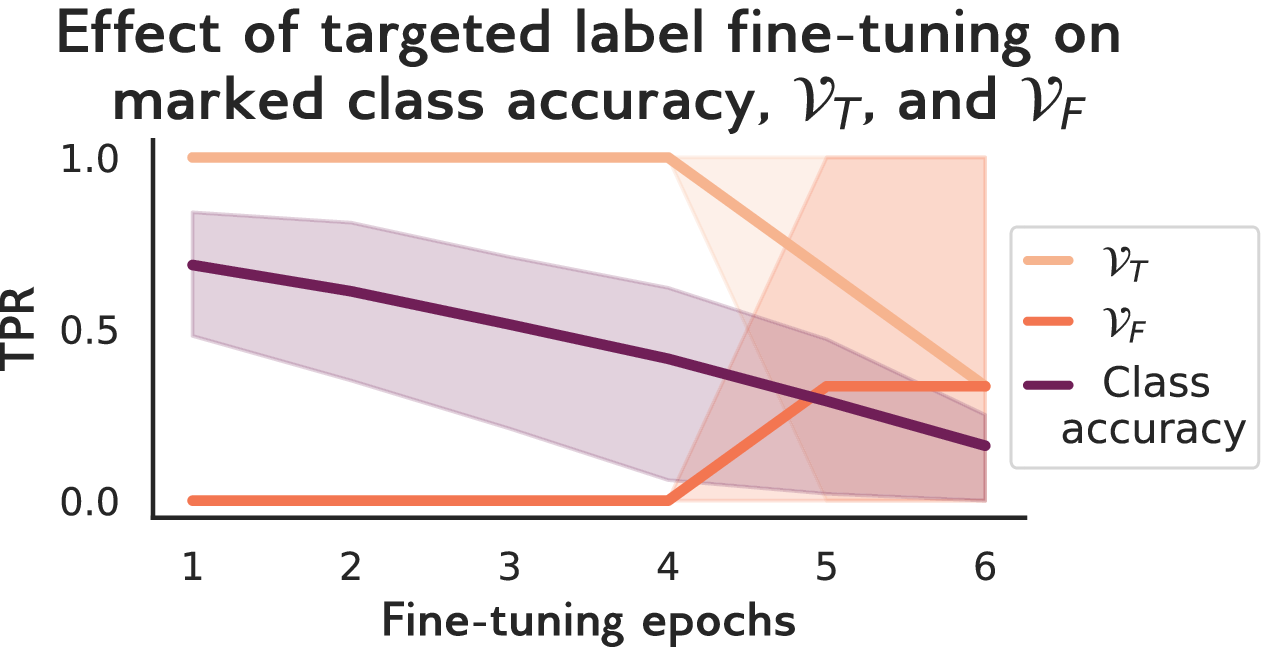}
      \vspace{-0.6cm}
      \caption{Retraining $\texttt{CIFAR100}$ tagged classes using $\texttt{Scrub}$ data drops accuracy faster than $\ver_T$.}
      \label{fig:retrain}
    \end{minipage}
    \end{center}
    \vspace{-0.6cm}
    \end{figure*}

\subsection{Spurious correlation detection}
\label{subsec:spurious}
\vspace{-0.1cm}

Isotope marking would be ineffective if \adv~could detect and filter out isotope images in $\train$. Existing literature has shown it is possible to detect spurious correlation in datasets~\cite{singla2021salient, moayeri2022spuriosity}. Since isotopes are inspired by the spurious correlation phenomenon, we test whether~\cite{singla2021salient}, a state-of-the-art spurious correlation detection method, can detect isoptoes in $\train$. \cite{singla2021salient} inspects feature maps produced by a trained $\model$ to see if spurious features caused $\model$'s classification decision. %The method then corrupts images with flagged spurious features from $\train$ and evaluates how much $\model$ relies on spurious features in its predictions.

Following~\cite{singla2021salient}, we run detection on \texttt{CIFAR100} models. Although~\cite{singla2021salient} assumes that the model is robustly trained, we omit this step, since the corresponding decrease in model accuracy~\cite{raghunathan2020understanding} hampers \adv's goal of training an effective model. We test the ``worst-case'' scenario for isotopes by computing feature maps for isotope images in $\train$ and manually inspecting whether isotope features are flagged in the list of top-$5$ most important features for the isotope class in $\model$, as reflected in the heatmaps. %This enables us to quickly check if~\cite{singla2021salient} flags isotope features. 
In reality, \adv~would not know which $\train$ images contain isotopes, so would have to inspect the top-$K$ activating features (depending on their threshold) for all $N$ classes. To understand the effect of mark visibility and frequency on detection, we vary $\alpha$ from $0.1$ to $0.5$ ($p=0.1$) and $p$ from $0.01$ to $0.3$ ($\alpha=0.5$). We use the single tagged class setting, which makes isotopes more likely to stand out and be detected as spurious features. 

\para{Results and cost.} For scenarios with smaller $p\le 0.2$ and $\alpha\le 0.4$, isotope features are not flagged (see Figure~\ref{fig:spurious}).  In the strongest cases (i.e. $\alpha=0.5$, $p \ge 0.2$), slight feature map shifts are observed, indicating that for these settings, this method may lead a model trainer to notice something ``odd'' about isotope images and possibly filter them. However, the $\alpha=0.5$, $p \ge 0.2$ setting is stronger than needed in practice for effective isotopes. Moreover, this method requires intense manual effort on the part of the model trainer to identify isotope images, making spurious correlation detection an impractical countermeasure. \emed{Outlier detection on the training dataset, a related method, also fails to detect spurious correlations (see Appendix~\ref{subsec:outlier} for details).}

\subsection{Feature inspection} 
\label{subsec:inspect}
\vspace{-0.1cm}

Inspecting $\model$'s features after training could enable detection of isotope-induced behaviors.  Since marks increase the probability of marked label $y_j$ for marked inputs, the feature space region associated with $y_j$ may exhibit isotope-specific behaviors.  Several defenses against backdoor attacks use feature
inspection to detect backdoors~\cite{tran2018spectral, chen2018detecting,schulth2022detecting, tang2021demon, peri2020deep,hayase2021spectre}. Here we adapt these to see if they can flag isotope behaviors. % if these defenses can detect isotopes.

We evaluate two feature inspection methods:
Spectral Signatures~\cite{tran2018spectral} and Activation
Clustering~\cite{chen2018detecting}. Both analyze the feature
representations of $\train$ elements in trained $\model$ and
run statistical tests to detect data that elicit unusual model
behaviors. Flagged data is removed from $\train$, and $\model$
retrained on the pruned dataset. We run both defenses using the
author-provided code adapted to our models. For Spectral Signatures,
we use the $95^{th}$ percentile as cutoff; for Activation Clustering,
we look for two clusters (e.g., ``clean'' and ``poison'') and use the
``smaller cluster" criterion, since there are fewer isotopes than
clean data.  We report average precision/recall in Table~\ref{tab:ac_ss}.

\begin{table}
\centering
\resizebox{0.3\textwidth}{!}{%
\begin{tabular}{ccc} 
  \toprule
                  & \textbf{\begin{tabular}[c]{@{}c@{}}Spectral\\ Signatures\end{tabular}} & \textbf{\begin{tabular}[c]{@{}c@{}}Activation\\ Clustering\end{tabular}} \\ \midrule
\textbf{Precision} & 0.004   & 0.018  \\
\textbf{Recall}    & 0.011    & 0.322    \\ \bottomrule  
\end{tabular}%
\vspace{-0.2cm}
}
\caption{Precision and recall of Spectral~\cite{tran2018spectral} and Clustering~\cite{chen2018detecting} on \texttt{CIFAR100} with $25$ marked classes.}
\label{tab:ac_ss}
\vspace{-0.6cm}
\end{table}

\para{Results and cost.} Both defenses have low precision and recall
in detecting isotope data. Less than $2\%$ of the data flagged by both
defenses is actually isotope data. Although Activation Clustering
has higher recall, detecting on average $32\%$ of isotope data, its
detection FPR is high ($36\%$). As
with spurious correlation detection, these methods have a nontrivial cost for \adv,
who must either manually filter the flagged data to find isotopes or
discard a large portion of $\train$. Neither defense detects
enough isotopes to significantly disrupt isotope detection.

\subsection{Adversarial augmentations}
\label{subsec:augment}
\vspace{-0.1cm}

If \adv~cannot find isotopes in $\train$ or $\model$, they can still
try to disrupt them.  One obvious way is to modify images in $\train$
during training.  Our experiments in \S\ref{sec:eval} employed common augmentation techniques during training, such as cropping, normalization, and rotation.  These did not
disrupt isotope performance, but we now test if more aggressive image
augmentation could prevent $\model$ from learning isotope features.

\para{Adding noise.} As a base case, \adv~could try to disrupt isotopes by adding Gaussian noise to $\train$ images before training. This could disrupt subtle features on images, potentially rendering marks ineffective. However, as Figure~\ref{fig:adapt_noise} shows, this is not the case. Adding noise with $\mu=0$ and varying $\sigma$ to $\train$ images (Fig.~\ref{fig:adapt_noise}) reduces $\model$ accuracy faster than $\ver_T$ or $\ver_{D_T}$.

\para{Adding additional marks.} A more aggressive tactic would be to add {\em more} marks to $\train$, to disrupt the learning of isotope marks. We assume \adv~adds marks to all images in $\train$, since they cannot know a priori which images have isotope marks. We use images from the $\texttt{GTSRB}$ dataset as \adv's additional marks and test their effect on isotope performance as $\alpha$ varies. 

As Figure~\ref{fig:adapt_newmark} shows, adding additional marks slowly degrades $\model$ and $\ver_{D_T}$ accuracy as $\alpha$ increases. However, it has a much stronger effect on $\ver_T$, which drops to 0 once the additional mark $\alpha' \ge 0.2$. This performance drop is likely because the new marks added are {\em extremely similar to} both the isotope and external marks used in $\ver$ (e.g. all are images blended into other images). When the model sees similar marks on all training images, isotope marks are no longer unique and are not learned as spurious correlations, confounding $\ver$. 

\para{Results and cost.} Adding noise imposes a significant {\em model accuracy cost} on \adv, as it causes to $\model$ accuracy degrade as or more quickly than $\ver_T$ and $\ver_{D_T}$. Since \adv~wants to train a highly accurate model, they would not use noise to disrupt isotopes. Although adding new marks drops $\ver_T$ once the additional marks have $\alpha' \ge 0.2$, model accuracy decreases by at least $5\%$ when $\alpha'=0.2$, which may unacceptable for \adv, depending on the setting. Regardless, we believe this countermeasure works better because of the similarity between the new marks and our isotope marks, making it more difficult to for isotope marks to act as spurious features. Future work broadening the set of mark options could mitigate this issue.

\vspace{-0.1cm}
\subsection{Reducing Granularity of Outputs}
\label{subsec:masking}

\adv~could try to prevent isotope detection by modifying $\model$'s outputs, since this could disrupt $\ver$. We consider two methods \adv~could employ: adding noise to $\model$'s logits or reducing the granularity of $\model$'s classification results.

\para{Add noise to $\model$ outputs.} Since $\ver$ uses differences in probabilities to detect isotopes, adding noise to $\model$'s outputs may obscure probability shifts and render $\ver$ ineffective. We test this by adding Gaussian noise with $\mu=0$ and varying
$\sigma$ to $\model$'s logits before computing the output probability vector. However, as Figure~\ref{fig:noisy_logits} shows, adding noise to $\model$'s logits degrades model accuracy before $\ver_T$ or $\ver_{D_T}$ decrease. Since \adv~incurs a high accuracy cost, this method is unusable.

\para{Return only top-$K$ predictions.} 
Our basic isotope detection algorithm assumes that $\model$ returns a
probability distribution over all classes. While this assumption
holds for many real-world ML APIs (Table~\ref{tab:api_settings} in Appendix~\ref{sec:appx_query}), $\model$ could respond to queries with less information (e.g. Face++ in Table~\ref{tab:api_settings}).  

To test isotope performance in this modified setting, we limit the
model's outputs to the top-$K$ ranked classes, $K \in \{2, 5, 10, 15, 20, 25, 50\}$ and compute the shift in the rank of the isotope class between $\mathbf{x_t}$ and $\mathbf{x_{t'}}$.  If the isotope class is not in the top $K$, we set its rank as $K+1$.  $\ver$ runs its t-test on the rank shifts, instead of probability shifts.  We report the average $\ver_T$ and $\ver_{D_T}$ accuracy for each $K$.

As Figure~\ref{fig:topk_results} shows, $\ver_T$ remains high in the rank-only setting, but $\ver_{D_T}$ decreases significantly.  Our explanation is that \emph{any correct} mark learned by $\model$, regardless of whether it is correct for a given class,
induces a change in $\model$'s probability, simply because it has been learned.  When raw probabilities are available, there is an
obvious distinction between the probability shift for true and false
marks for a class.  When only the top-$K$ outputs are available, there is not enough signal determine this.  While this drop in $\ver_{D_T}$ in the top-$K$ setting is unfortunate, recall that an individual
user $U$ only knows their mark and thus cannot run $\ver_D$.  Therefore, top-$K$ outputs are sufficient for detecting the mark, the user's primary goal.

\para{Results and cost.} Adding noise to $\model$'s logits directly decreases model accuracy and imposes a significant cost on the model trainer.  The cost of restricting to only the top-$K$ outputs is subtler. Unlike other countermeasures, this technique would, in many settings, reduce the model's utility for users. Furthermore, limiting outputs to only ranks provides only ``security by obscurity'', and could likely be overcome by more advanced isotope distinguishing methods~\cite{choquette2021label, li2021membership}.

\vspace{-0.1cm}
\subsection{Targeted Fine-tuning}
\label{subsec:retrain}
\vspace{-0.1cm}

Finally, a motivated adversary can fine-tune their model with unrelated data to make $\model$ ``forget'' isotope-related features. In adapting to new data, $\model$ might hold onto core features of the original class but forget spurious features like isotopes. To test this, we resize, relabel, and normalize $\texttt{Scrub}$ images
to serve as fine-tuning data for tagged labels in $\texttt{CIFAR100}$ (see Figure~\ref{fig:retrain}).  Marked class accuracy
degrades much faster than $\ver_T$, making targeted retraining costly
and ineffective.

\section{Limitations and Future Work}
\label{sec:discussion}

There are a number of limitations to our current work. First, most of our experiments use visibility level $\alpha=0.4$, which can leave visible marks on images. We made the tradeoff for this higher $\alpha$ because it means we can detect isotopes with near-perfect accuracy when isotopes only make up $10\%$ of a model class. This might be an acceptable cost for privacy conscious users, but can easily be adjusted per user preferences. 

Second, we did not explore isotope efficacy in other scenarios, e.g. enterprise scale models with millions of classes, or lower $p$ values below 0.1, for scenarios where many users contribute data to a common class. Third, our approach can be affected by model trainers who only offer limited classification output (e.g. only top-K results), or those who are willing to sacrifice their own model accuracy to evade isotope detection (\S\ref{subsec:augment}). Finally, despite our best efforts to study a range of adaptive attacks, it is possible our system can be circumvented by future countermeasures.

There are also several directions to extend and improve this work. First, the isotope marks we evaluate -- ImageNet images blended into other images -- introduce large feature disturbances into images. There is clearly ample room for work that explores alternative approaches with significantly less visual impact, e.g. spurious correlations that do not require a mask over the full image. Second, we need to better understand how isotopes (and other data provenance tools) behave in a continual learning setting, as is used in many commercial ML models today~\cite{kumar2020understanding, nahar2022collaboration}. While results in \S\ref{sec:countermeasures} show that retraining with orthogonal data does not cause a model to forget isotope features, long-term retraining of models with in-distribution data could over time cause forgetting of isotope features, since they are not ``core'' class features. 
\newpage

\bibliographystyle{acm}
{\footnotesize
\bibliography{main}}

\appendix
%\normalfont
\section{Appendix}

\begin{table*}[t]
  \centering
  \resizebox{0.92\textwidth}{!}{
  \begin{tabular}{crccc}
  \toprule
  \textbf{Task} & \textbf{Classes} & \textbf{Model} & \textbf{Loss} & \textbf{Training setting}  \\ \midrule
  GTSRB   & 43  & Simple & Cross-entropy & Adam(lr=$0.0001$, epochs=20, batch size=512)\\
  CIFAR100 & 100 & ResNet18 & Cross-entropy & SGD(lr=0.5, scheduler=step, epochs=72, batch size=512) \\
  PubFig  &  65 & SphereFace (pretrained) & Angle & Adam(lr=$0.001$, epochs=25, batch size=128) \\
  Scrub & 530 & ResNet50 & Focal & Adam(lr=$0.001$, scheduler=cyclic, epochs=16, batch size=128)  \\ 
  ImageNet  & 1000 & ResNet50 & Cross-entropy & Adam(lr=$1.7$, scheduler=step, epochs=18, batch size=512)\\ \bottomrule
  \end{tabular}%
  }
  \vspace{-0.1cm}
  \caption{Model training details for each task.}
  \label{tab:training_settings}
\end{table*}

\subsection{Model Architectures and Training}
\label{sec:appendix_train}
 We use different model architectures and training procedures for each task. The training settings for each dataset are in Table~\ref{tab:training_settings}. For most tasks, models are trained from scratch. The exception is $\texttt{PubFig}$, due to its small size, which we train via transfer learning from models pre-trained on the CASIA-Webface dataset~\cite{yi2014learning}. All experiments are run on our local servers using 1 NVIDIA GPU. For \texttt{CIFAR100}, we use the ffcv library to expedite training~\cite{leclerc2022ffcv}. 

\subsection{$\ver$ baseline performance and boosting}
\label{sec:appendix_boost}

In our experiments, we run $\ver$ using {\em boosting}, i.e. multiple runs of the t-test, in order to minimize randomness. Here, we explore the effect of $Q$, the number of boosting rounds, on the TPR/FPR of $\ver$. The goal is to use the minimum number of boosting rounds that produce a stable $\ver$ performance, to minimize the cost of verification. We also explore the baseline TPR/FPR for $\ver$ when it is run on $t'_1$ and $t'_2$, two external marks. $\ver$ should have roughly random performance (TPR $\approx$ FPR) in this setting. 

\begin{figure*}[t]
    \begin{center}
    \begin{minipage}[c]{0.45\linewidth}
    \centering
    \includegraphics[width=0.8\textwidth]{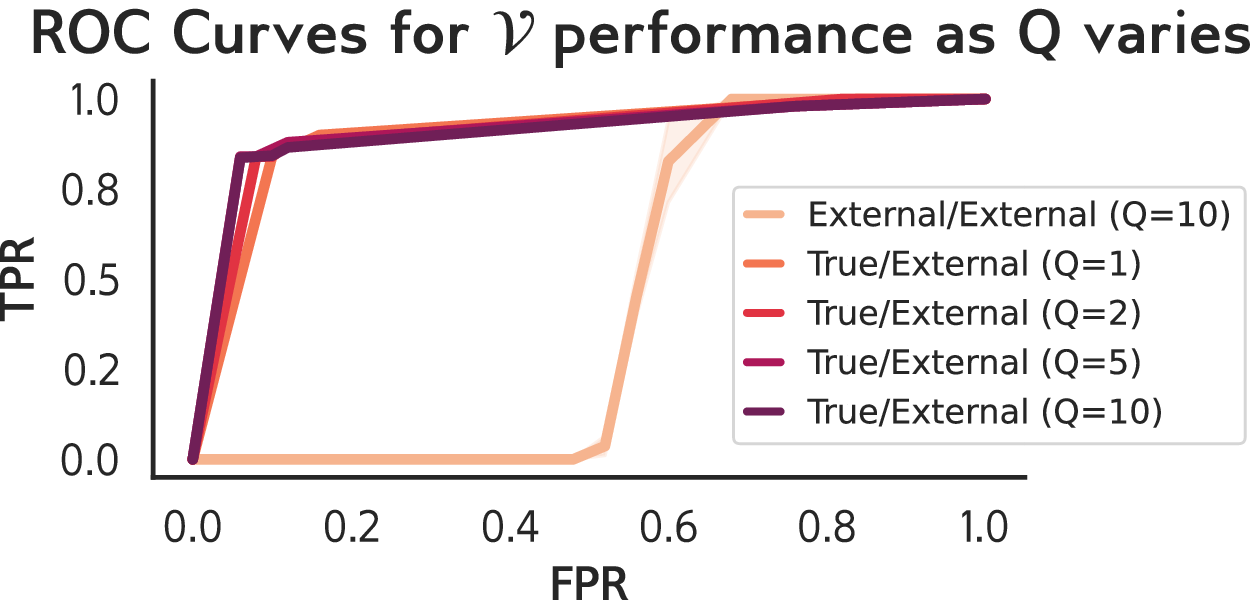}
    \caption{Comparison of $\ver$ performance for different $Q$ values and on paired external marks.}
    \label{fig:boost_compare}
  \end{minipage}\hfill
  \begin{minipage}[c]{0.45\linewidth} 
    \centering
    %\resizebox{1\textwidth}{!}{%
    \begin{tabular}{clll}
        \toprule
              & $\mathbf{\texttt{set1}}$ & $\mathbf{\texttt{set2}}$ & $\mathbf{\texttt{set3}}$ \\ \midrule
    $\%$ with match &  $95\%$  & $94\%$   &  $80\%$ \\
    Avg. mark match rank & 1.0   &   1.0  &  4.8  \\ 
    Avg. true match rank &   1.0   &   1.0   &   -  \\ \bottomrule
    \end{tabular}%
    \vspace{0.1cm}
    %}%%
    \captionof{table}{Results from isotope detection in Amazon Rekognition. For $\texttt{set1}$ and $\texttt{set2}$, the true match is always the top match. For unenrolled isotope images ($\texttt{3}$), isotope images with the same mark appear in the top $5$ hits.}
    \vspace{-0.2cm}
    \label{tab:rekognition}
    \end{minipage}
  \end{center}
  \vspace{-0.5cm}
\end{figure*}

To test this, we evaluate a $\texttt{CIFAR100}$ model with $30$ marked classes, $\alpha=0.5$, $p=0.1$. We run $\ver$ using different $Q$ values on both true/external mark pairs (as typically used in $\ver$) and external/external mark pairs (for baseline performance calibration). As Figure~\ref{fig:boost_compare} shows, $\ver$'s performance slightly improves when going from $1$ to $5$ boosting rounds, but increasing from $5$ to $10$ does not significantly improve performance. Thus, in our \S\ref{sec:eval}-\S\ref{sec:countermeasures} experiments, we use $Q=5$. As expected, results for the external/external $\ver$ tests are random, even when $Q=10$.

\begin{figure*}[t]
  \centering \includegraphics[width=0.99\textwidth]{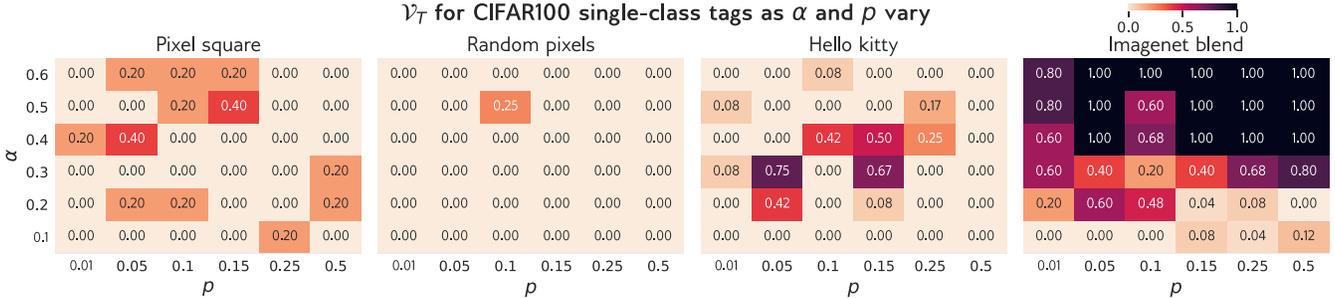}
  \vspace{-0.1cm}
  \caption{Average $\ver_T$ values for different marks in a $\texttt{CIFAR100}$ model. Marks that introduce stronger features into images (like Hello Kitty and Imagenet Blend) perform much better. For a discussion of different mark performance, see Appendix \ref{sec:appendix_other_tags}}
  \label{fig:different_marks_cifar}
  \vspace{-0.3cm}
  \end{figure*}
  
\subsection{Performance of different marks}
\label{sec:appendix_other_tags}
Using the parameters and training settings described in \S\ref{subsec:method}, we train \texttt{CIFAR100} models with isotopes created using the four marks shown in Figure~\ref{fig:mark_types}. To explore how mark settings impact performance, we vary $\alpha$ from $0.1$ to $0.6$ (see Figure~\ref{fig:mark_visibility}) and $p$ from $0.01$ (e.g. 1\% of data marked) to $0.5$. Figure~\ref{fig:different_marks_cifar} reports the average $\ver_T$ for each setting. Overall, we find only Imagenet blend marks are consistently detectable. This indicates that marks with more unique and diverse features make marks better isotope candidates, and once such a mark is visible and frequent enough in a user's data, it can be detected. 

The pixels square, random pixels, and Hello Kitty marks {\em can} induce probability shifts for classes to which they are added, as illustrated in Figure~\ref{fig:example_spurious}. However, these marks do not produce {\em strong enough} probability shifts to be robust to the false positives test $\ver$ runs \textemdash e.g. comparing the true mark to some external mark. This false positives test is necessary to make isotopes practically useful, and when we employ this, we find that the pixels square, random pixels, and Hello Kitty marks are less effective. Thus, we use the Imagenet blend mark in the rest of our experiments.

\subsection{Isotopes in facial recognition engines}
\label{sec:rekognition_appendix}

Testing isotopes in commercial FR systems requires some modifications
to the detection algorithm.  Today, these systems work by matching query
images to a reference database via \emph{feature space similarity}, as
opposed to directly applying a trained ML model.  Standard approaches
involve measuring $L_2$ similarity between the query and reference
images in the feature space of a trained DNN~\cite{arcface, cosface,
magface, facenet}. Reference images that are similar (or identical)
to a queried image are returned as the ``top match''.  We leverage this
fact to detect isotopes in commercial FR databases.

We run experiments on Amazon Rekognition, a popular facial recognition
engine that allows users to build a reference image database and submit
new images to the database via an API~\cite{rekognition}. Rekognition
does not disclose how images are processed in their system, what DNN is
used to produce features, or how feature space matching is performed.

We enroll $100$ people from the $\texttt{Scrub}$ dataset in a Rekognition
database using $100$ images/person. We select $10$ $\texttt{Scrub}$
classes for isotope testing, $5$ men and $5$ women ($4$ Black, $6$
Caucasian). For each, we enroll $5$ different images with the same mark
($\texttt{set1}$) and $5$ images that are identical but have different
marks ($\texttt{set2}$). At test time, we set the confidence threshold
(the minimum similarity for a reference image to be returned as a match)
at 95 and query $\texttt{set1}$, $\texttt{set2}$, and $\texttt{set3}$,
which contains new images with the $\texttt{set1}$ mark. All marks are
ImageNet blend marks with $\alpha=0.4$. In Table~\ref{tab:rekognition},
we report the proportion of images for which any isotope match was
returned, the average rank (1 = best) of the first isotope image in the
match set, and the average rank of the true match for $\texttt{set1}$,
$\texttt{set2}$.

As Table~\ref{tab:rekognition} shows, $\texttt{set1}$ and $\texttt{set2}$
always have the true enrolled image as their top match, i.e., we perfectly
detect isotopes in the database.  Interestingly, $\texttt{set3}$ images,
which have the same mark as $\texttt{set1}$ but are not enrolled, have
an isotope image appear in the top $5$ matches on average, even though
isotopes are only $10\%$ of the enrolled set, i.e., a marked query image
often draws out \emph{other} isotopes with the same mark enrolled in
the database.

\para{Discussion.} Isotope detection described above exploits the fact
that FR engines are very good at matching identical images.  Thus, if a
user knows what images they posted online and where, they can determine
if a particular source was included in an FR database by querying the
corresponding FR engine with an image from that source.  Isotopes are
not strictly necessary for this sort of auditing: if an exact image
is in the reference database, it will typically be the top match.
Isotopes can still be useful for users to quickly ``sort'' which site
the images came from, perhaps by posting identical images with different
marks on different sites.

\subsection{Outlier detection countermeasure}
\label{subsec:outlier}

\begin{figure}[!h]
  \begin{center}
\begin{minipage}[c]{0.9\linewidth}
    \centering \includegraphics[width=0.8\textwidth]{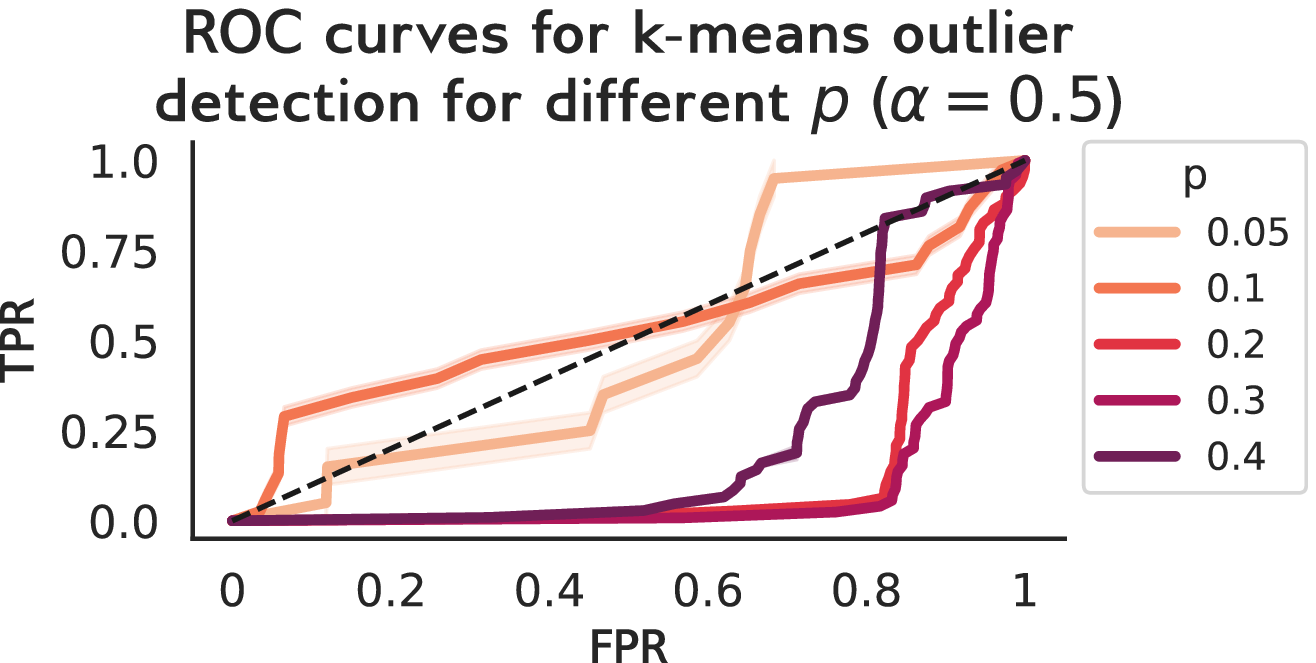}
    % \vspace{-0.2cm}
    \caption{Outlier detection with fixed $\alpha$ and varying $p$. As $p$ decreases, isotopes become rarer and outlier detection performance slightly improves.}
    \label{fig:outlier_detection}
  \end{minipage}\hfill
  \begin{minipage}[c]{0.9\linewidth}
    \vspace{0.2cm}
    \centering \includegraphics[width=0.8\textwidth]{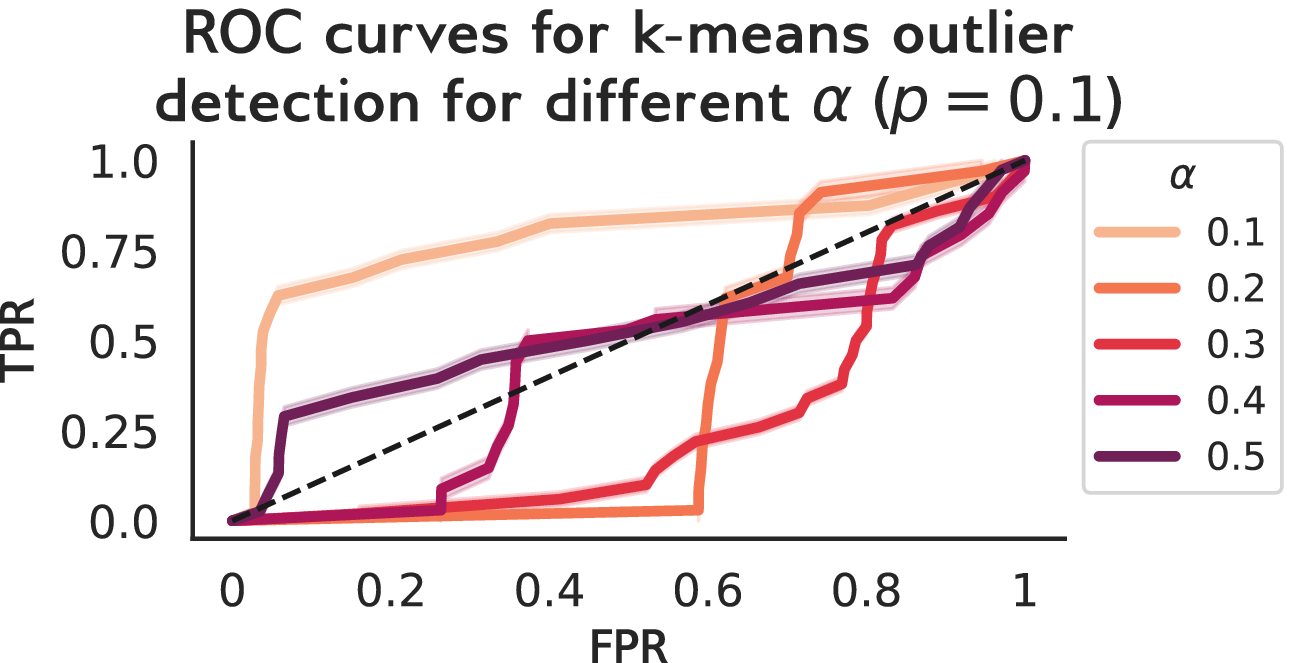}
    % \vspace{-0.2cm}
    \caption{ROC curves for outlier detection with fixed $p$ and varying $\alpha$. The method works best for $\alpha=0.1$, but $\ver_T$ is low for this $\alpha$/$p$ (\S\ref{subsec:single_mark}).}
    \label{fig:outlier_detection2}
  \end{minipage}\hfill
\end{center}
\vspace{-0.6cm}
\end{figure}

Outlier detection could enable \adv~to discover marked
images in the training dataset and remove them before training
$\model$. To test the efficacy of this countermeasure, we run an outlier detection method that is based on k-nearest neighbors~\cite{original_knn}. We pass $\train$ through a model pre-trained on a similar domain to create feature
representations and cluster the representations into $N$ classes,
where $N$ is the number of classes in $\train$. Finally, we run outlier detection
on these clusters while varying the outlier threshold to compare the TPR
(i.e. isotope data flagged as outlier) and FPR at different thresholds. We assume that \adv~looks for isotope outliers for each label/cluster.

Since we test on $\texttt{CIFAR100}$ models, we use a pre-trained $\texttt{Imagenet}$ model to produce the feature representation.  We evaluate in the single mark
setting, since this represents the \emph{most optimistic} scenario for
the model trainer: with only one label marked, isotope images are more
likely to stand out and be flagged as outliers. To understand the effect
of mark visibility and mark frequency on detection efficacy, we vary $\alpha$ from $0.1$ to $0.5$ ($p=0.1$) and $p$ from $0.01$ to $0.3$ ($\alpha=0.5$). 

\para{Results and cost.} As Figures~\ref{fig:outlier_detection} and~\ref{fig:outlier_detection2} show, when $\alpha$ is larger or $p$ is smaller, isotope images are easier to flag as outliers, and the AUC for outlier detection increases. Outlier detection performs well
when $\alpha=0.1$ and $p=0.1$, but $\ver$ accuracy is low for
these parameters, making them unlikely to be used in practice (see
Figure~\ref{fig:different_marks_cifar}). Overall, KNN-based outlier
detection detects isotope outliers only at high false positive rates,
necessitating either additional filtering to find the true positives or
throwing out a large chunk of unmarked data.  This is a nontrivial cost, as both acquiring new data and manually inspecting existing data are time- and resource-intensive.  More advanced outlier detection
may reduce the FPR, and we leave this as future work.

\subsection{Query Outputs in Real World Systems}
\label{sec:appx_query}

\begin{table}[!h]
  \centering
  \resizebox{0.5\textwidth}{!}{%
  \begin{tabular}{cclc}
  \toprule
  \textbf{Task}                          & \textbf{Service}   & \textbf{Query Output}                  & \textbf{Reference} \\ \midrule
  \multirow{3}{*}{Face recognition}      & Rekognition & All labels above threshold &    \cite{aws_confidence}            \\
   &  Azure         & All labels above threshold &  \cite{azure_confidence} \\
   & Face++                  & Up to top 5 matches                          & \cite{faceplusplus_topk}  \\ \midrule
  \multirow{2}{*}{Object classification} & Apple ML kit       & All matches above threshold &    \cite{apple_ml_confidence}                \\
   & Google ML Kit & Flexible, default = top 5  &  \cite{automl_confidence}\\ \bottomrule
  \end{tabular}%
  }
  \vspace{-0.1cm}
  \caption{Prediction outputs returned by different ML services. Most
  services return all labels that match the input with more than a certain
  ``confidence'' threshold level, set by the user performing the query.}
  \label{tab:api_settings}
  % \vspace{-0.4cm}
  \end{table}

Table~\ref{tab:api_settings} provides examples of query outputs returned by real-world MLaaS providers. Most systems by default return any matches (for facial recognition) or labels (in a classification setting) above a given confidence threshold. Users interacting with the MLaaS API can vary this threshold in their queries to obtain more (or fewer) results from the model. The one exception to this rule is Face++, a platform for building custom facial recognition engines, which will return at most the top $5$ query results.

\end{document}